# Place-level urban-rural indices for the United States from 1930 to 2018

Johannes H. Uhl[1,2*], Lori M. Hunter[1], Stefan Leyk[1,3], Dylan S. Connor[4], Jeremiah J. Nieves[1,5], Cyrus Hester[1], Catherine Talbot[1], and Myron Gutmann[1,6]

[1]Institute of Behavioral Science, University of Colorado Boulder, Boulder, CO-80309, USA.
[2]Cooperative Institute for Research in Environmental Sciences, University of Colorado Boulder, Boulder, CO-80309, USA.
[3]Department of Geography, University of Colorado Boulder, Boulder, CO-80309, USA.
[4]School of Geographical Sciences & Urban Planning, Arizona State University, Tempe, AZ-85287, USA.
[5]Department of Geography and Planning, University of Liverpool, Liverpool, L69 7ZT, UK.
[6]Department of History, University of Colorado Boulder, Boulder, CO-80309, USA.

*Corresponding author: Johannes H. Uhl (johannes.uhl@colorado.edu), Institute of Behavioral Science, University of Colorado Boulder, 483 UCB, Boulder, CO 80309-0483, USA.

**Abstract**: Rural-urban classifications are essential for analyzing geographic, demographic, environmental, and social processes across the rural-urban continuum. Most existing classifications are, however, only available at relatively aggregated spatial scales, such as at the county scale in the United States. The absence of rurality or urbanness measures at high spatial resolution poses significant problems when the process of interest is highly localized, as with the incorporation of rural towns and villages into encroaching metropolitan areas. Moreover, existing rural-urban classifications are often inconsistent over time, or require complex, multi-source input data (e.g., remote sensing observations or road network data), thus, prohibiting the longitudinal analysis of rural-urban dynamics. Here, we develop a set of distance- and spatial-network-based methods for consistently estimating the remoteness and rurality of places at fine spatial resolution, over long periods of time. We demonstrate the utility of our approach by constructing indices of urbanness for 30,000 places in the United States from 1930 to 2018 and further test the plausibility of our results against a variety of evaluation datasets. We call these indices the *place-level urban-rural index (PLURAL)* and make the resulting datasets publicly available (https://doi.org/10.3886/E162941) so that other researchers can conduct long-term, fine-grained analyses of urban and rural change. In addition, due to the simplistic nature of the input data, these methods can be generalized to other time periods or regions of the world, particularly to data-scarce environments.

**Keywords**: Rural-urban continuum, urban gradient, long-term population dynamics, human settlements, spatial demography, spatial network analysis.

1. Introduction

Over the twentieth century, the substantial growth of towns and cities across the United States profoundly reshaped the nation's population distribution. The share of the US population living outside of urban areas fell from roughly 60 percent in 1900 to less than 20 percent today, and many communities that were once rural were absorbed into cities through urban expansion. Data constraints have, however, limited our understanding of how this process has unfolded at fine spatial scales while also limiting understanding of current conditions in rural communities, particularly those with smaller populations.

In order to study rural and urban processes, researchers have already generated many indices, classifications, and typologies of rurality, based on a wide range of data (Nelson et al. 2021). However, existing measures of the rural-urban continuum face a combination of three challenges: a) they are generally derived from county-level data which is often too coarse a scale to describe the population dynamics of rural places; b) they lack temporal consistency which prohibits longitudinal analysis; and/or c) they are constructed based on measures of urbanized land rather than population size. In this article, we evaluate existing characterizations of the rural-urban continuum and propose new classification approaches that address the three issues above. Particularly for rural settings, we contend that a better understanding of sub-county units – especially the spaces where rural dwellers focus their daily, collective activities – is essential for research, planning, and the development of place-relevant policies and programs. The classification approach presented here advances efforts to address this important gap.

**1.1. Existing Classifications of the Rural-Urban Continuum**

Many rural-urban processes and their associated effects play out at the sub-county scale (e.g., places, towns, and villages), and therefore, are masked by county scale analyses. The failure to observe the spatial heterogeneity of population processes produces substantial biases such as when data reflecting coarse units (e.g., county) are used to characterize the finer units within them (e.g., places) (Hunter et al., *forthcoming*).

For a wide variety of reasons, researchers often use the county as an analytical unit (e.g. Curtis et al. 2020; Machado et al. 2021). The county is often a policy-relevant choice representing local stakeholders within multi-level governance (Homsey, Liu, and Warner 2019). In addition, a vast amount of information is available at the county scale including sociodemographic characteristics (e.g., U.S. Census 2019a), health outcomes and behaviors (e.g., CDC 2021), mortality (e.g., Curtin and Spencer 2021), and many indices reflecting for example, the rural-urban continuum (e.g., Golding & Winkler 2020). Specific to the rural-urban continuum, indices have been used in a wide variety of studies including analysis of spatial variation in income inequality (Thiede et al. 2020), mortality (Brooks, Mueller, and Thiede 2020), political polarization (Scala & Johnson 2017), and myriad other social processes (e.g., Pender et al. 2019, Johnson & Lichter 2020, Lichter & Johnson 2020, Lichter & Johnson 2021). General profiles of rural America are also often county-based (e.g., Cromartie et al. 2020).

Examples of existing rural-urban classifications in the US include the commonly used *rural-urban continuum codes (RUCC)* created by the US Department of Agriculture's (USDA) Economic Research Service (ERS). The RUCC identify nine categories, i.e., three metro and six nonmetropolitan county designations, with metropolitan counties further disaggregated by the encompassing metro area's population size (McGranahan et al 1986, Butler 1990; Fig. 1a). Nonmetropolitan counties are further classified by their degree of urbanization and adjacency to a metro area. Golding & Winkler (2020) refined the RUCC to distinguish explicitly between urban cores and their exurbs and suburbs, resulting in the *rural-urban gradient (RUG)* (Fig. 1b).

The USDA ERS also produces the *rural-urban commuting area (RUCA) codes* (ERS 2013; also available by the ZIP code area). The RUCA codes make use of the U.S. Census urbanized areas and urban core designations (U.S. Census Bureau 2020), in combination with census-tract level commuting flow estimates. Combined, the RUCA groups census tracts into 10 classes of commuting levels (Fig. 1c). Another related measure is the *USDA urban influence codes (UIC)* (Ghelfi & Parker 1997) which yields nine different classes based on the population of the county's largest city rather than an aggregated urban population as in RUCC (Fig. 1d).

The National Center for Health Statistics (NCHS) released a classification called the *urban-rural classification scheme (URCS)* (Ingram & Franco 2014) based on metropolitan and non-metropolitan county classification in combination with population thresholds, identifying six county designations (Fig. 1e). Moreover, a continuous classification scheme is provided by the *index of relative rurality (IRR)* (Waldorf 2006 and Waldorf & Kim 2018; Fig. 1f), a county-level index based on population size, density, road network distance, and built-up areas. As the IRR method is independent from administrative or census-defined boundaries, the underlying framework can be applied to finer-grained spatial units as well (Waldorf & Kim 2015).

Finally, there are the *Frontier and Remote (FAR) Area Codes* available at the ZIP code level (Cromartie & Nulph 2015), which provide four classes of remoteness, and are derived from travel times and population estimates. Herein, the FAR codes are used for cross-comparison and are discussed in detail in Section 2.4.1. See Waldorf & Kim (2015) and National Academy of Sciences (2016) for reviews of these various classification approaches.[1]

Despite different data sources and methodological approaches, these six classification schemes show generally high levels of correlation (Fig. 1g). The lowest correlation is between IRR and RUCA (Pearson = 0.59) likely driven by the different data sources (i.e., built-up density vs. commuting patterns). Moreover, the continuous IRR exhibits a positive association with the rank-based metrics, which is, on average, strongest and almost linear between IRR and URCS; it shows the least nuanced trend between IRR and the tract-level RUCA (Fig. 1h).

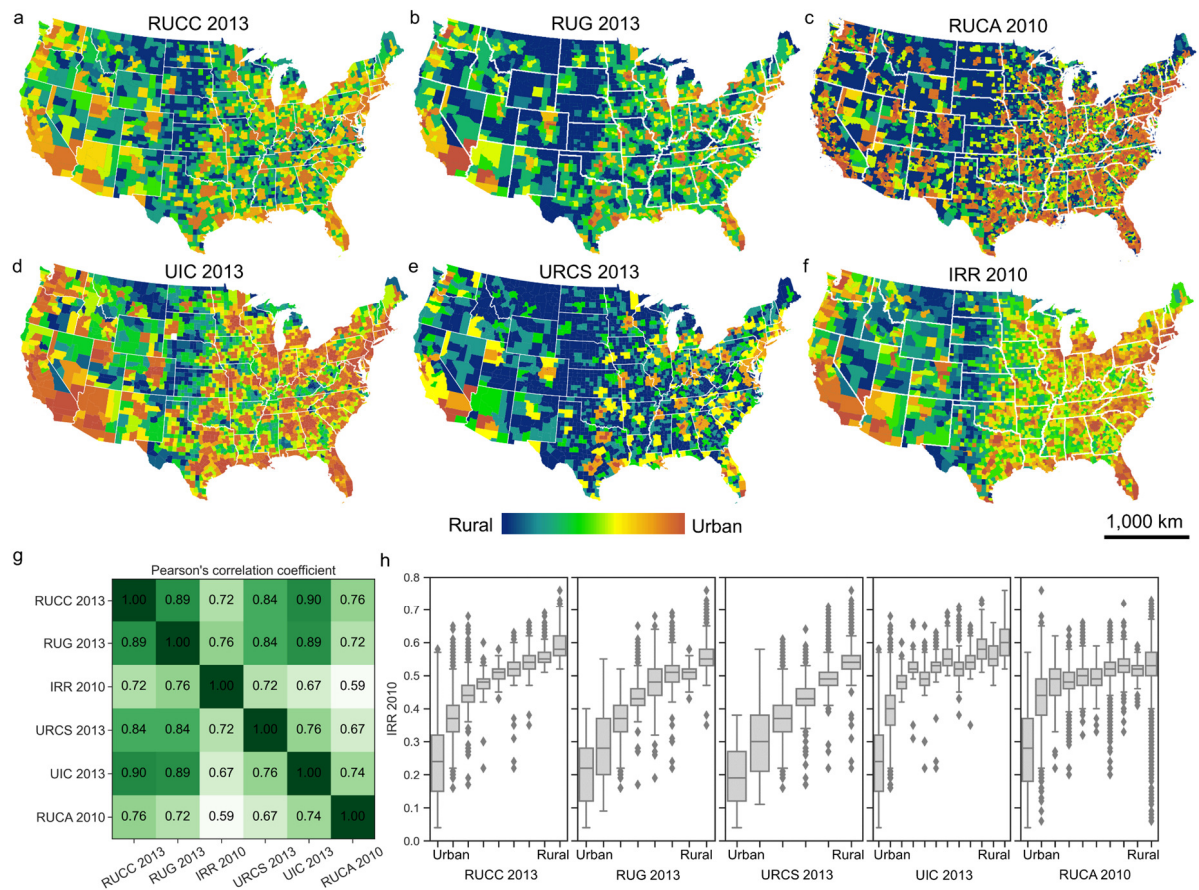

**Figure 1. Visual-analytical comparison of existing rural-urban classification schemes in the U.S.: (a) USDA rural-urban continuum codes (RUCC) in 2013 (b) the Rural-Urban Gradient (RUG) 2013, (c) USDA rural-urban commuting areas (RUCA) in 2010 at the tract-level, (d) USDA urban influence codes (UIC) 2013, (e) 2013 NCHS Urban–Rural Classification Scheme (URCS), and (f) the Index of Relative Rurality (IRR) in 2010. Panel (g) shows cross-correlations between the different rural-urban classifications, computed at the tract-level for RUCA, and at the county-level otherwise. Panel (h) illustrates the variations of the continuous IRR against the rank-based, categorical classification schemes.**

---

[1] Many countries provide individual delineations of urban and rural areas (Workman & McPherson 2021), often relying on census-based information. At a global scale, researchers typically rely on classifications of the rural-urban continuum derived from remotely sensed earth observation data in combination with population estimates, such as GRUMP (Balk et al. 2005), degree of urbanization (Dijkstra & Poelman 2014), GHSL-SMOD (Florczyk et al. 2019). However, these methods are confined to recent decades, of relatively coarse spatial resolution, and represent a land perspective more than a population view.

## 1.2. Limitations of Existing Categorizations of the Rural-Urban Continuum

As described in the Introduction, existing measures of the rural-urban continuum typically face at least one of three challenges of relevance to rural populations, each described below.

*County-scale data are relatively coarse.* Indices at the county-scale face important challenges. First, the Modifiable Areal Unit Problem (MAUP) is well-established in the social and spatial sciences and refers to the fact that relationships observed at aggregate units often do not hold at finer scales ( Openshaw and Taylor 1979; Flowerdew et al. 2001). We contend that while counties provide great utility in understanding rural change (e.g., Cosby et al. 2019; Johnson & Lichter 2019; Monnat 2020), their coarseness can lead to misleading conclusions regarding sub-county units and processes (Homsy et al. 2019).

County-scale analyses can also face the Uncertain Geographic Context Problem (UGCoP) which refers to the challenge whereby relevant conclusions depend on analyses at spatial units corresponding to the true and relevant geographic contexts experienced by individuals (Kwan 2012). Here we can look to a prominent body of recent rural-related work that examines how the characteristics of places shape individual-level processes including social mobility (Chetty et al. 2014; Connor & Storper 2020), racial inequality (Manduca & Sampson 2019), health, and voting (e.g., Shah et al. 2020; Sachdeva et al. 2021). Indeed, individual lives are typically lived in communities, places that influence life trajectories and places to which meaning can be ascribed. Such meaning furthers a sense of belonging and the development of place-based identities (e.g., Sack 1997, Manzo & Devine-Wright 2013, Armstrong and Stedman 2019). Place identity appears especially strong among rural dwellers (Lewicka 2005, Anton and Lawrence 2010) and, while today's rural America is ever-changing, recent work confirms that, in general, rural residents remain deeply tied to place and hold strong commitments to community (Ulrich-Schad and Duncan 2018). As compared to ZIP codes, places have the advantage of being representative of a meaningful social unit. ZIP codes were created to increase the efficiency of mail delivery and can be especially problematic as an analytical unit in rural areas (Grubesic 2008). "Places", as defined by the U.S. Census provide functions for a concentration of people. They are locally recognized, independent of other places, and can be either incorporated places – defined by criteria within their respective states – or census-designated places (CDPs), which are not incorporated and lack a municipal government (U.S. Census 2008). As such, places are of both practical and social importance in the lives of rural dwellers (Federal Register 2008). An important constraint of place as a unit of analysis, however, is that it neglects consideration of residents outside of place boundaries. We contend that, as collectives, places serve nearby residents as well as those within specific boundaries.

*Lack of temporal consistency.* County boundaries change over time, as do the designations of metropolitan and non-metropolitan counties. As a result, classifications such as the RUCC suffer from temporal inconsistencies caused by changes in methodology, and by changing units that cannot be compared between different points in time. Of course, place boundaries also change across time and, as a result, we do not explicitly engage place-based boundaries in the approach articulated below. Instead, we use place population data and incorporate a broader, more general representation of spatial extent based on distance to other places (details below).

*Based on limited and hard-to-acquire data.* Some of the more complex indices (e.g., the IRR, FAR) are grounded in data reflecting road networks or built-up areas, information that does not typically offer substantial historical coverage. Such data are more challenging to acquire compared to Census data.

## 1.3. A Place-Level Urban-Rural Index

Overall, efforts to analyze demographic processes across the rural-urban continuum at the place level and over time have been impeded by the lack of spatially fine-grained and temporally consistent indicators of rural and urban places. Because "urbanness" or "rurality" are multivariate processes that evade simple definition, we use the concept of "remoteness" to continuously measure the urban-to-rural spectrum. We consider places as "remote" (i.e., rural) if they have relatively small populations and are surrounded by other small places. Maximum remoteness is achieved if these surrounding, small places are also very distant. We refer to "non-remote" (i.e., urban) places if they are relatively large in population and/or surrounded by other large places.

We propose two methods to derive measures of remoteness of places in the U.S. (and possibly elsewhere) at fine spatial granularity. These methods generate consistent classifications of rural and urban places over long periods of time by implementing simplified characteristics commonly used to define rural-urban classes (e.g., size, distance, local importance and spatial relationships between populated places). Specifically, the first approach is based on population size of places and the weighted (Euclidean) distances to other places of different size categories. While this approach is computationally efficient and can be implemented as a raster-based approach, it may overly generalize local spatial configurations of populated places, and thus, ignore valuable information regarding the local importance of a place. Thus, we propose a second approach, based on a spatial network, that adopts concepts from landscape ecology and network analysis to model remoteness in a more spatially explicit manner. We call the proposed indices the ***place-level urban-rural indices (PLURAL)***. We name the raster-based index PLURAL-1, and the spatial network-based index PLURAL-2. Both approaches rely on the same data input, which are solely derived from public-domain data and allow for the derivation of various (combined) distance and population-based attributes to model remoteness based on different perspectives.

Such refined indicators enable new possibilities for understanding pressing urban and rural issues, such as disparities in regional development, infrastructure, and social and economic well-being. Until now, analyses of these issues have often been constrained to relatively coarse scales of analysis. Focusing on the spatial distributions of the remoteness index also allows for direct examination of the changing nature of urban and rural places. Analysis of the remoteness indices over time provides unprecedented insight into the development history and urbanization of the United States and our reliance on publicly available data in the creation of these fine-scale indices provides an accessible and flexible option for scholars and policymakers, particularly those concerned with issues affecting small places and other data scarce environments.

Herein, we describe the derivation of the PLURAL indices and their underlying data for the conterminous US (CONUS), as well as a range of cross-comparisons and plausibility analyses (Section 2). We then demonstrate the applicability of the PLURAL for modelling the long-term dynamics of the rural-urban continuum by measuring place-level remoteness in the CONUS for each decade, from 1930 to 2018 (Section 3), and assess the plausibility of the calculated place-level rural-urban classifications by comparing against a range of external data sources (Section 4). We conclude with a critical evaluation and discuss future directions (Section 5). The indices for the time period from 1930 to 2018 are publicly available for download as tabular and spatial datasets at https://doi.org/10.3886/E162941.

## 2. Data and Methods

In this section, we describe the input data and the derivation of the PLURAL-1 index based on gridded surfaces (i.e., raster-based approach) (Section 2.2). We then describe PLURAL-2, a spatial network-based remoteness modelling approach that explicitly accounts for local spatial relationships between populated places by adopting concepts from network analysis and landscape ecology (Section 2.3). We also introduce the data sources and strategies used for cross-comparison and plausibility analysis of the results (Section 2.4). An overview of the presented approaches is shown in Figure 2.

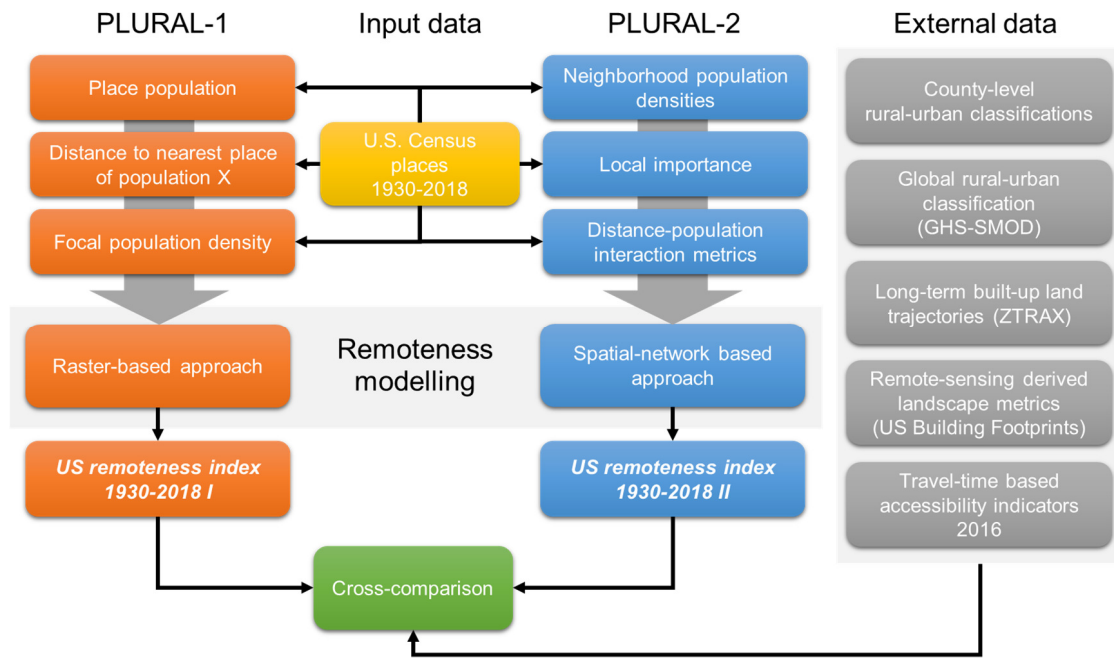

**Figure 2. Flow diagram of the presented methods.**

## 2.1. Source data and preprocessing

In this work, we use US census place population counts from 1970, 1980, 1990, 2000, and 2010, obtained from NHGIS[2] (Manson et al. 2020), containing both, incorporated places and census-designated places (U.S. Census Bureau 1994). From the same data source, we obtained point-based census place locations for each decade between 1930 and 2010, as well as for 2018, and census place polygons for 2010. Moreover, we use census place populations (I.e., 5-year population estimates 2014-2018) from the American Community Survey (ACS). Importantly, we digitized decadal place populations for 1930 to 1970, obtained from counts published in the 1940 and 1960 decennial reports (Tables 5 and 8, respectively; U.S. Census Bureau 1942, 1964) (see Fig. 3a-c for some examples). We then joined the tabular data to the spatial data and used these integrated datasets as base data for all subsequent data processing and analyses. In total, we obtained 213,827 place locations, across all years (from 15,641 places in 1930, to 28,814 places in 2018), attributed with their population counts.

## 2.2. Derivation of the raster-based remoteness index (PLURAL-1)

Using place-level population estimates (1930-2018), provided for discrete geospatial locations, we design a method to model the remoteness of places across the region of interest (e.g., the US). The remoteness index of a place is computed based on the size $s$ (i.e., population) of the place of interest and the distance between that place and the nearest places of varying size categories (10,000-20,000, 20,000-50,000, 50,000-100,000, 100,000-250,000 and more than 250,000 people, herein referred to as population categories $pc$, Fig. 3d-h). These or similar categories are used in numerous studies (e.g., Angel et al. 2011, Cromartie & Nulph 2015, Nelson et al. 2019). Moreover, a focal population density $pd$ within a radius of 10km is used (Fig. 3i) in order to characterize the population distributions in the place neighborhood, and to reduce the sensitivity of the index to arbitrary partitioning of places (e.g., neighborhoods in large cities are typically recorded as individual places). We chose a radius of 10km since based on some initial experiments, a circle with a 20km diameter is likely to meaningfully aggregate the individual neighborhoods of a large city, without overly aggregating dispersed rural places.

---

[2] https://www.nhgis.org/

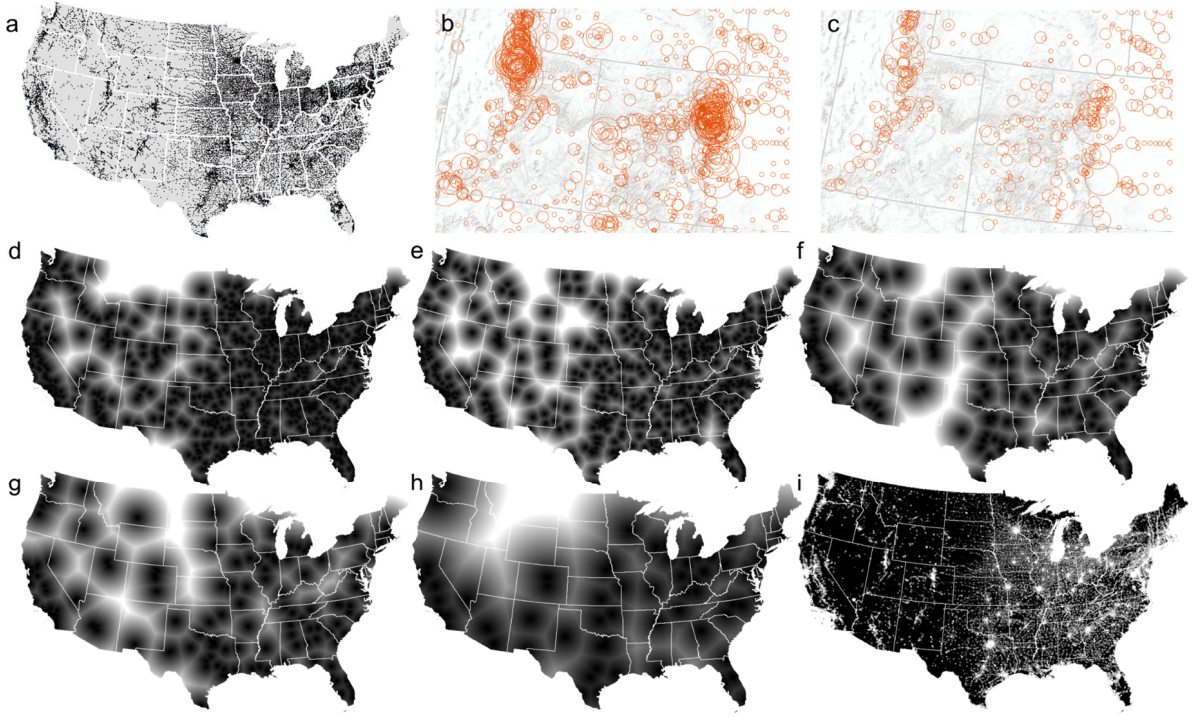

**Figure 3.** Source data and raster-based processing steps: (a) US census places in 2018, (b) place-level population for Colorado and Utah in 2018 and (c) in 1930. Panels (d) to (h) show the distance surfaces to the nearest place of population (d) 10,000 to 20,000 (e) 20,000 to 50,000, (f) 50,000 to 100,000, (g) 100,000 to 250,000, and (h) >250,000 derived from the place population estimates in 2018. Panel (i) shows the focal population density surface derived from the 2018 place populations.

The place-level urban-rural index **PLURAL-1** for a given place $p$, in a given year $t$ can be derived as the weighted average of the inverse of the population size $s_{p,t}$, of place $p$ in year $t$, and the distance measures $D_{pc,p,t}$, (in meters) to the nearest place of population category $pc$ (Equation 1). All measures are log-transformed to achieve a uniformly distributed index despite skewed distributions of population and population density (and potentially skewed distributions of distance measures due to the presence of extremely remote places):

$$\overline{PLURAL}_{p,t} = \left[ w_s \cdot \log\left(\frac{s_{p,max}}{s_{p,t}}\right) + w_{pd} \cdot \log\left(\frac{pd_{max}}{pd_{p,t}}\right) \sum_{pc=1}^{5} w_{pc} \cdot \log(D_{pc,p,t}) \right] \quad (1)$$

$$\text{with } w_{p,t} + w_{pd,t} + \sum_{pc=1}^{5} w_{pc} = 1.0 \quad (2)$$

where $w_s$ is the weight for place population, $w_{pd}$ is the focal population density weight, and $w_{pc}$ are weights for the distance measures to different population categories $pc$ to allow for adjusting the influence of local versus regional population centers. The constants $s_{p,max}$ and $pd_{max}$ are global maximum values of place population and focal population density, respectively, and can either be derived from the data distribution or chosen based on domain knowledge. Herein, we use a maximum place population of $s_{p,max} = 10,000,000$ and a maximum focal population density of $pd_{max} = 15,000$ people / km². By log-dividing the upper bounds of population and population density by the place-level values, $\log\left(\frac{s_{p,max}}{s_{p,t}}\right)$ and $\log\left(\frac{pd_{max}}{pd_{p,t}}\right)$, respectively, we obtain measures that yield low values for large, densely populated places.

This results in a total of seven **remoteness indicators** (based on population, population density, and five distance measures). We propose four different weighting schemes:

- **Equal weights:** All remoteness indicators are weighted equally.
- **Place-centric:** Place population and focal population density share 50% of the weight, and each distance-based component receives a weight of 0.1.
- **Place-centric + metro focus:** Place population and focal population density share 50% of the weight, and the weights for the distance components are constructed such that distance to large (metro) areas has the highest weight.
- **Metro focus:** Place population and population density receive low weights, and the distance-based components receive higher weights, with the distance to places >250,000 receiving the highest weight.

Herein we focus on the equal weights scenario for simplicity, but refer readers to Table A1 and Figures A1 and A2, for greater detail of the influence of individual weighting schemes. The final raster-based index is then calculated by scaling the raw index measures $\overline{PLURAL}_{p,t}$ into the range [0,1]:

$$PLURAL_{p,t} = \frac{\overline{RI}_{p,t} - \min(\overline{RI}_t)}{\max(\overline{RI}_t) - \min(\overline{RI}_t)} \quad (3)$$

This computation yields values close to 0 for large places near other (large and/or small) places, and values close to 1 for small places, remote from other places. By approximating each place by a discrete point location (i.e., the place polygon centroid), rather than using its areal extent, and by modelling the distances between places using Euclidean rather than road network distances, our approach is highly versatile and generalizable to data-scarce environments and (early) time periods, as retrospective areal place extents and multi-temporal road network data are rarely available for these periods.

**2.3. Modelling remoteness based on spatial networks**

The raster-based approach presented in Section 2.2 is computationally inexpensive. That is, population density and distance-based components can be derived from distance grids easily in commonly used GIS environments. However, this approach may ignore the local, spatial configuration of populated places, which may contain important information regarding the local importance of a place. Thus, we use concepts from network analysis and landscape ecology to provide a second modelling approach. Such methods and metrics have been applied to human settlement modelling based on remote-sensing derived patches of built-up land (Esch et al. 2014) or for analyzing global land cover patterns (Nowosad & Stepinski 2018). Using a network to describe the spatial configuration of the point-based places allows for the derivation of topology-based, and thus, density-independent metrics. This is particularly important as the population and settlement density across the United States varies considerably across space and time. Similarly, utilizing local landscape metrics enables the quantification of the localized, place-centric configuration of neighborhood place populations. Moreover, this network-based approach allows for a joint, place-centric assessment of neighboring places, where the raster-based approach (Section 2.2) only considers the nearest places of each population category only without taking into account the whole spectrum of the spatial context (e.g., the *n*-th nearest place) which may contribute to the rurality of a given place as well.

2.3.1. Establishing place-level spatial networks

Populated places may be given as discrete point locations (see Fig. 3b,c), or, typically for recent points in time, as areal objects (Fig. 4a). In this case, place locations from 1980 onwards are given as polygons, and, prior to that, as discrete point locations attributed with their place population. For consistency, we converted place polygons into discrete locations by using their centroid coordinates, and generated Thiessen polygons (Voronoi 1908, Thiessen & Alter 1911) based on these discrete locations (Fig. 4b). Topological relationships between the Thiessen polygons allowed us to construct spatial networks for different levels of neighborhood cardinality which can be understood as varying scales of spatial context.

The concept of neighborhood cardinalities is used to complement the Euclidean distance-based measures and allows us to identify neighborhood relationships between places independent from spatial density variations. For example, neighbors of cardinality 1 are direct neighbors of a node (i.e., connected by an edge), and a cardinality of 2 includes the neighbors of cardinality 1 of the cardinality 1 neighbors of a given place, etc. This is relevant when having a consistent but locally flexible method that can be applied to the densely populated Northeast as well as the sparsely populated regions in the Southwest of the U.S. These networks consist of nodes (i.e., place locations) and edges (i.e., connections between neighboring places), shown in Fig. 4c for the neighborhood of cardinality 1 (i.e., connecting places whose Thiessen polygons share a common boundary). These topological relationships enable the efficient identification of different neighborhood levels for each place (e.g., Fig. 4d).

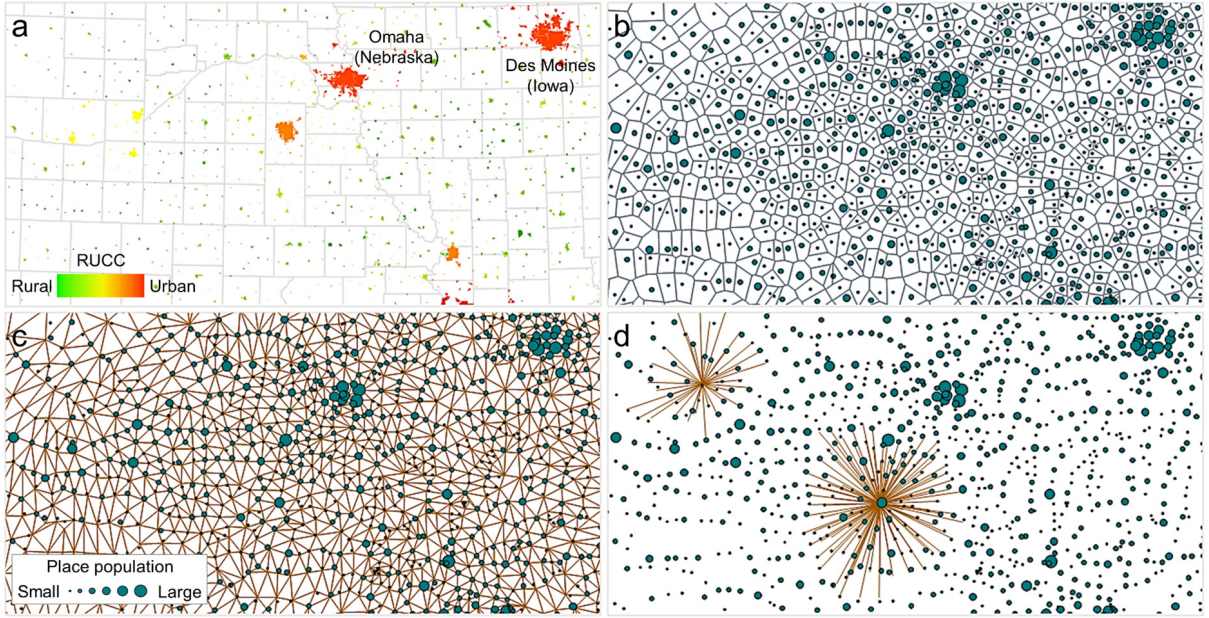

**Figure 4. Illustrating the spatial network generation: (a) US census places in 2010, color-coded by 2013 USDA county-level rural-urban continuum codes, (b) Thiessen polygonization, (c) spatial network for neighbors of cardinality 1, (d) exemplary neighborhoods of cardinality 3 (top left) and cardinality 5 (center).**

2.3.2.  Spatial-network based remoteness indicators

Similar to the focal population density metric, based on a fixed focal radius used in the raster-based approach (Section 2.2), we modeled ***population density*** in local, topology-based neighborhoods: For each place $p$, we identify the neighboring places of a given cardinality ($p_c$) and calculated the total population $s_c$ in the neighborhood. To calculate the approximate population density, we use a square of size nxn with n being the largest occurring distance $dist_{max}$ between $p$ or any of the neighboring places $p_{c1}$. Thus, the neighborhood population density (***NPD***) for any cardinality $c$ of place $p$ and year $t$ can be calculated as:

$$NPD_{p,t,c} = \frac{[s_{p,t} + s_{c,t}]}{dist_{max}^2} \quad (4)$$

To capture local population density at different levels of spatial context around each place, we calculated the ***NPD*** for the cardinalities 1, 2 and 3. Moreover, we adopted a measure of ***local significance***, which was proposed by Esch et al. (2014). The local significance ***LS*** can be calculated for each edge of a spatial network, based on the length of the edge $d$ (i.e., the Euclidean distance between neighboring places) and the size $s$ (i.e., population) of the nodes (places ***i*** and ***j***) connected by the edge, as:

$$LS_{i,j} = \frac{[s_i \times s_j]}{|d_{i,j}|^2} \quad (5)$$

Similar to Esch et al. (2014), we calculated the median local significance $MLS_C$ for each place $p$ based on the values obtained for each edge connected to the node that represents the place $p$. Accordingly, we calculated this metric for each place within neighborhoods of cardinality 1, 2 and 3. The MLS yields high values if a place is large and located near other large places, and yields small values if the neighboring places of a given small place are small and distant.

In the raster-based remoteness modelling approach (Section 2.2) we used the distance to the nearest place of a given population category $D_{PC}$ as a measure of remoteness. However, such a metric ignores the spatial configuration of places below the chosen population threshold that could potentially be located between the place under study and the nearest place of a given population range. For example, a place (A) located 50km from the nearest city > 50,000 inhabitants (B) would receive the same value for $D_{PC}$, regardless if the area between A and B is a completely uninhabited (scenario 1) or if that area contains many small places below the lowest population threshold, and outside the focal window size used for the focal population density calculation (scenario 2). The degree of remoteness of place A should be higher in case of scenario 1 than in scenario 2.

Motivated by this shortcoming of the $D_{PC}$ metrics, we adopted the concept of proximity and isolation metrics, a subgroup of landscape metrics commonly used in landscape ecology and habitat fragmentation analysis (e.g., Bender et al. 2003) to quantify the degree of subdivision of a landscape or of the isolation of specific components (e.g., land cover classes) within a landscape. More specifically, we adopt the concepts of the "degree of landscape division" metric (DIV), proposed by Jäger (2000) and distance-weighted landscape variables (see Miguet et al. 2017). Jäger (2000) defines the DIV as the area under the curve when sorting patches in a given landscape by their patch area, as a measure of the graininess of a landscape (McGarigal 1995). Based on these, we designed a place-centric, distance-based metric quantifying the relationships of neighboring place populations and their distances to the "focal" place (i.e., the place under study) in a single metric. Our method identifies the neighbors of a given focal place, either using a Euclidean distance or a topology-based neighborhood criterion, and sorts the neighboring places, including the focal place itself, ascendingly by their distances to the focal place. Then, the cumulative population curve is calculated *over the distance-sorted places*, and finally, the area under the cumulative population curve is obtained. This area under the curve (AUC) represents a metric characterizing the spatial distribution of populated places in dependence of the distance to the focal place. We call this metric the **distance-based neighborhood population index (DNPI).**

For illustration and as a proof of concept, we show the cumulative, distance-sorted population curves for 2010 census places in the U.S., located in urban counties (RUCC 1, Fig. 5a), in peri-urban counties (RUCC 5, Fig. 5b) and rural counties (RUCC 9, Fig. 5c). If the focal place is near a large place, the cumulative population curve will increase sharply and yield a large area under the curve (Fig. 5a). Conversely, if a place is small and its neighbors are small too, the cumulative population curve increases only slowly, yielding a small AUC. The visual assessment (Fig. 5) clearly confirms that the DNPI responds to the degree of urbanness. In order to make this metric comparable between different focal places, the maximum distance $d_{MAX}$ needs to be specified, as well as a maximum value for the cumulative population $CP_{MAX}$, and the curves need to be scaled by $d_{MAX}$ in x-direction and by $CP_{MAX}$ in y-direction, respectively, so that it is normalized into the range [0,1]. If the curve exceeds $CP_{MAX}$ before $d_{MAX}$ is reached, the cumulative population curve is "trimmed" to 1.0. Thus, the maximum possible AUC is 1.0 for a place of population >= $CP_{MAX}$. Thus, we calculated the DNPI within the neighborhood of cardinality 3, and for a range of distance-population combinations. Table 1 summarizes the network-based metrics of remoteness used herein.

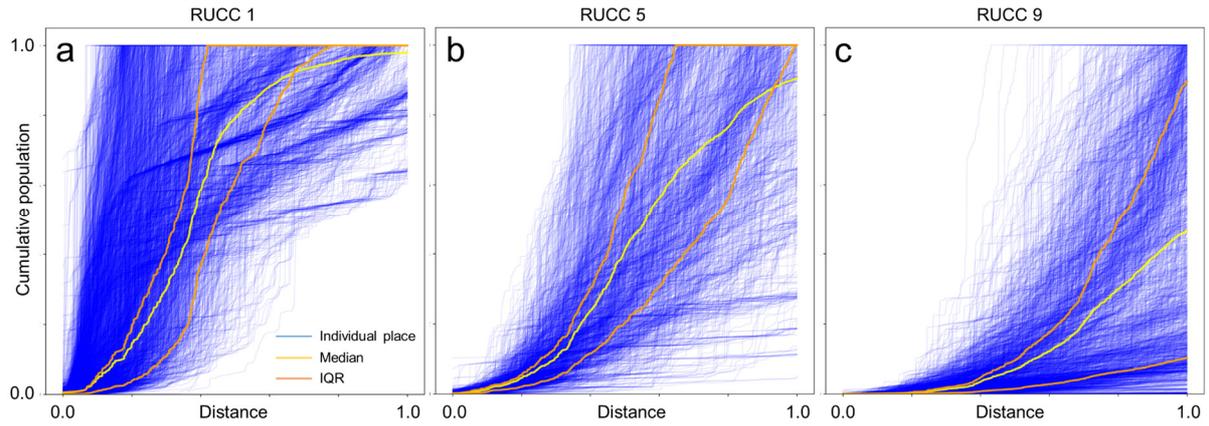

**Figure 5.** Illustrating the concept of the distance-based neighborhood population index (DNPI). Cumulative population curves over the neighboring places, sorted by their distance, provide an area under the curve which characterizes the spatial configuration of place populations with respect to a "focal" place. Shown are the curves based on U.S. census places 2010 within counties of (a) RUCC 1 (urban), (b) RUCC 5 (peri-urban), and (c) RUCC 9 (most rural) in 2013. The stopping criteria used are a maximum distance of 500km, and a maximum cumulative population value of 1,000,000.

**Table 1. Overview of the network-based remoteness metrics.**

| Metric | Reference neighborhood | Description |
|---|---|---|
| **Population** | | |
| $POP_{Place}$ | Place | Population of the place (i.e., node) |
| $NPD_1$ | Cardinality 1 | Population density of places in neighborhood of cardinality 1, referred to the squared maximum distance between places in neighborhood |
| $NPD_2$ | Cardinality 2 | Population of places in neighborhood of cardinality <=2, referred to the squared maximum distance between places in neighborhood |
| $NPD_3$ | Cardinality 3 | Population of places in neighborhood of cardinality<= 3, referred to the squared maximum distance between places in neighborhood |
| **Local significance** | | |
| $MLS_1$ | Cardinality 1 | Median local significance (MLS) of edges connecting each place with its neighbors of cardinality 1 |
| $MLS_2$ | Cardinality 2 | MLS of edges connecting each place with its neighbors of cardinality <=2 |
| $MLS_3$ | Cardinality 3 | MLS of edges connecting each place with its neighbors of cardinality <=3 |
| **Distance-based neighborhood population indices** | | |
| $DNPI_{C3}$ | Neighborhood of cardinality <= 3 | DNPI (i.e., AUC of cumulative population of neighbors, sorted by their distance), within neighbohood of c=3. |
| $DNPI_{250km, 500k}$ | 250km | DNPI within radius of 250km, or until reaching a cumulative population of 500,000 |
| $DNPI_{500km, 1,000k}$ | 500km | DNPI within radius of 500km, or until reaching a cumulative population of 1,000,000 |
| $DNPI_{MAXPOP}$ | $d_{MAX}$ | DNPI until reaching a cumulative population equal to the population of the largest place in the analyzed distribution, or the maximum occuring distance $d_{MAX}$. |

The maps of these 11 metrics for the 2018 places are shown in Fig. A4. For each of these metrics, we calculated the ranks in a descending order (i.e., the lowest magnitude receives the highest rank), and use these remoteness indicators as input for different weighted averages. We implemented the following weighting schemes:

- **Equal weights:** All remoteness indicators are weighted equally.
- **Population focus:** Place population and NPD metrics share 50% of the weight, all other metrics receive equal weights.
- **DNPI focus:** DNPI based metrics receive 50% of the weight, all other metrics receive equal weights.
- **Significance focus:** MLS-based metrics receive 50% of the weight, all other metrics receive equal weights.

Herein we focus on the equal weights scenario to generate the PLURAL-2 for each U.S. census place in the years 1930 - 2018, but refer to Tables A2 and Figure A3 which discuss the influence of individual weighting schemes in greater detail. Importantly, we calculated both indices using a) annual scaling to the range [0,1], and b) scaling across all years 1930-2018, in order to generate temporally comparable remoteness indices for longitudinal studies. Note that the network-based indices PLURAL-2 are based on a ranking strategy, which makes the indices not directly comparable over time, and thus should not be used for longitudinal analysis. Fig. A5 presents the frequency distributions of both modelling approaches, all weighting schemes and all years.

**2.4. Spatio-temporal evaluation of the remoteness indices**

We compared our raster-based (PLURAL-1) and spatial network based (PLURAL-2) remoteness indices to a variety of external, independent datasets that are coherent or correlated to the concept of remoteness, or that are assumed to follow systematic patterns across the rural-urban continuum. Note that a rigorous validation is not possible, as the concept of remoteness is not a perfect substitute for rurality or urbanness, and existing data are not generally measured at the scale of places. Further, definitions of remoteness likely have many truths, much as there are many definitions of urban and rural. More specifically, we compared the PLURAL-1 and PLURAL-2 indices against each other (Section 3.2), to quantify the effects of the different modelling approaches, and against the discussed county-level classifications (Section 2.4.1). Additionally, we used the GHS-SMOD data (Florczyk et al. 2019) to test our approaches against a global, remote-sensing derived urban-rural classification (Section 2.4.2), and compared our multi-temporal results to historical settlement trends derived from the Historical Settlement Data Compilation for the U.S. (HISDAC-US, Leyk & Uhl 2018, Uhl et al. 2021) (Section 2.4.3) and against landscape metrics derived from Microsoft's building footprint data (Microsoft 2018) (Section 2.4.4). Lastly, we compare our created place-level remoteness indices against travel-time based accessibility indicators (Nelson et al. 2019) (Section 2.4.5).

2.4.1. County- and ZIP code level rural-urban classifications

The county-level rural-urban classifications shown in Fig. 1 provide valuable baseline models on the rural-urban continuum in the U.S. We employed these to evaluate whether the created remoteness indices exhibit similar trends, measured by the correlation between our place-level indices and the rural-urban designation assigned to the county containing each place; and assessed the outliers in the place-level distributions of the PLURAL indices within strata of county-level rural-urban designations (Section 3.3). Moreover, we compared our place-level results to the FAR area codes at the ZIP code area level. There are four FAR area codes, temporally referenced to 2010, identifying ZIP code areas with populations living more than specific travel time thresholds from urban areas of specific population thresholds (see Cromartie & Nulph 2015 for details). As opposed to the county-level classifications, the FAR classes are not mutually exclusive. Thus, we use receiver-operator-characteristic (ROC) analysis (Green & Swets 1966) to test whether there are thresholds that can be applied to our remoteness indices that yield high levels of agreement when comparing to each of four FAR classes (Section 3.4). To join ZIP code area FAR designations to the places, we applied a spatial join based on 2010 place polygon centroids to the 2010 ZIP code areas (U.S. Census Bureau 2019b).

2.4.2. Global rural-urban classifications (GHS-SMOD)

Since the county-level urban-rural classifications discussed in Section 1 were designed exclusively for the U.S. and exhibit a relatively large temporal gap to the most recent set of U.S. census places (i.e., ACS population estimates from 2018), we decided to compute a county-level urbanness index for the U.S. based on the recently released, globally available Settlement Model (SMOD), compiled by the Global Human Settlement (GHS) project (Pesaresi et al. 2013). The GHS-SMOD is based on population data and built-up areas derived from Landsat observations (Pesaresi et al. 2016) and implements the REGIO/EUROSTAT taxonomy (Dijkstra & Poelman 2014) for defining classes of urbanness (Florczyk et al. 2019). GHS-SMOD is available globally at a spatial resolution of 1km. The GHS-SMOD classifies the Earth into 1km grid cells of seven density-based levels of urbanness, ranging from "very low density" (class 11) to "urban center" (class 30) (Pesaresi & Freire 2016). We calculated the area proportions of each class per county, and computed a weighted average per county, based on these area proportions, giving weight 1 to the "very low density" class, and weight 7 to the "urban center". We then scaled the resulting county-level scores into the range of [0,1] (Fig. 6a) and assessed the distributions of our place-level indices within strata of county-level, SMOD-based urbanness estimates for consistency and plausibility (Section 3.3).

2.4.3. Long-term trends of built-up land trajectories

Although the temporal patterns in our remoteness indices directly reflect spatial population change over time, we would also expect these changes to be correlated with other processes related to urbanization, such as urban and rural building patterns. Using spatially explicit data on built-up areas for most of the U.S., at fine spatial granularity (i.e., a grid of 250m × 250m resolution) from the HISDAC-US, we constructed place-level trajectories of total built-up area and the number of buildings over time for each place where HISDAC-US data is available, for each decade from 1930 to 2010, as well as for 2015. We then calculated the correlations of these metrics with the PLURAL indices over time, with the expectation that the PLURAL indices (indicating remoteness) would be negatively correlated with new building (indicating urban expansion). Examples of the place-level built-up areas, over time and for three approximate levels of rurality, as indicated by the county-level RUCC, are shown in Fig. 6b, illustrating the different growth trajectories across the rural-urban gradient. Similarly, Fig. 6d illustrates how building density increases in peri-urban areas, over the long term, while staying relatively stable in scattered, rural settlements, strengthening our hypothesis of different built-up land trajectories along the rural-urban gradient (Section 3.5).

2.4.4. Contemporary settlement and landscape metrics derived from high-resolution remote sensing data

While the built-up areas as provided by the HISDAC-US are derived from cadastral property data and may suffer from low levels of spatial accuracy in rural areas, we also used Microsoft's building footprint data (Microsoft 2018), reflecting the state of built-up areas in approximately 2016, derived from high-resolution remote sensing imagery, at high levels of accuracy (Uhl et al. 2021). We rasterized these building footprint data to fine, CONUS-wide spatial grids of 250x250m (Uhl et al., in preparation). From these gridded surfaces (i.e., indicating the number of buildings, and the total building footprint area per grid cell), we computed commonly used settlement metrics, for each place, such as the number of buildings, total built-up area, average building area (measuring built-up intensity), and landscape metrics, such as the average area of contiguous patches of built-up land, number of patches, largest patch index (measuring spatial segregation), landscape division index, and patch cohesion index (measuring the segregation and connectedness of built-up land) (McGarigal 2015). We then assessed the correlations of these settlement and landscape metrics with our remoteness indices in 2018, motivated by previous work reporting strong relationships between the size and structure of built-up land and the rural-urban gradient (Luck & Wu 2002, Vizzari 2011, Vizzari 2013). More specifically, we used the bounding box of each 2018 place polygon, buffered by 1km in all directions, as a focal window in which the landscape metrics were computed, and assessed the correlation between these landscape metrics and the PLURAL indices (Section 3.6). By doing so, we were able to not only characterize the size, shape, and structure of the built-up areas representing each place, but also the unincorporated land in proximity to the places. Some examples of the extracted built-up areas per place are shown in Fig 6c, illustrating the difference in size and structure of built-up areas across the rural-urban continuum.

2.4.5. Travel-time based accessibility indicators

The accessibility indicators used herein are derived from globally available travel time estimates (Nelson 2019). These travel time estimates are based on road network data from OpenStreetMap and ancillary data on travel times, on land cover and terrain characteristics, as well as urban areas derived from the Global Human Settlement Layer (see Nelson et al 2019 and Weiss et al. 2018 for detailed method descriptions). These indicators are available as global, gridded surfaces at a spatial resolution of 30 arc-seconds, indicating the travel time from each grid cell to the nearest city of a specific population range, approximately in 2015 (see Fig. 6e for some examples). We used the surfaces associated with population ranges comparable to the population ranges used for the distance-based PLURAL index (Section 2.2), i.e., travel times to the nearest place of 10,000-20,000, 20,000-50,000, and 50,000-100,000 inhabitants (Fig. 6e), and extracted the travel time for each place centroid of our 2018 places dataset. We then qualitatively assessed the relationship between these travel times and our remoteness indices, and computed the theoretical travel speed based on the (road-network based) travel time given from the accessibility indicators, and the Euclidean distances used to construct the distance-based PLURAL index in 2018. We assessed the plausibility of these theoretical travel speed estimates in order to quantify the bias of using Euclidean distances instead of road network distances in our remoteness models (Section 3.7).

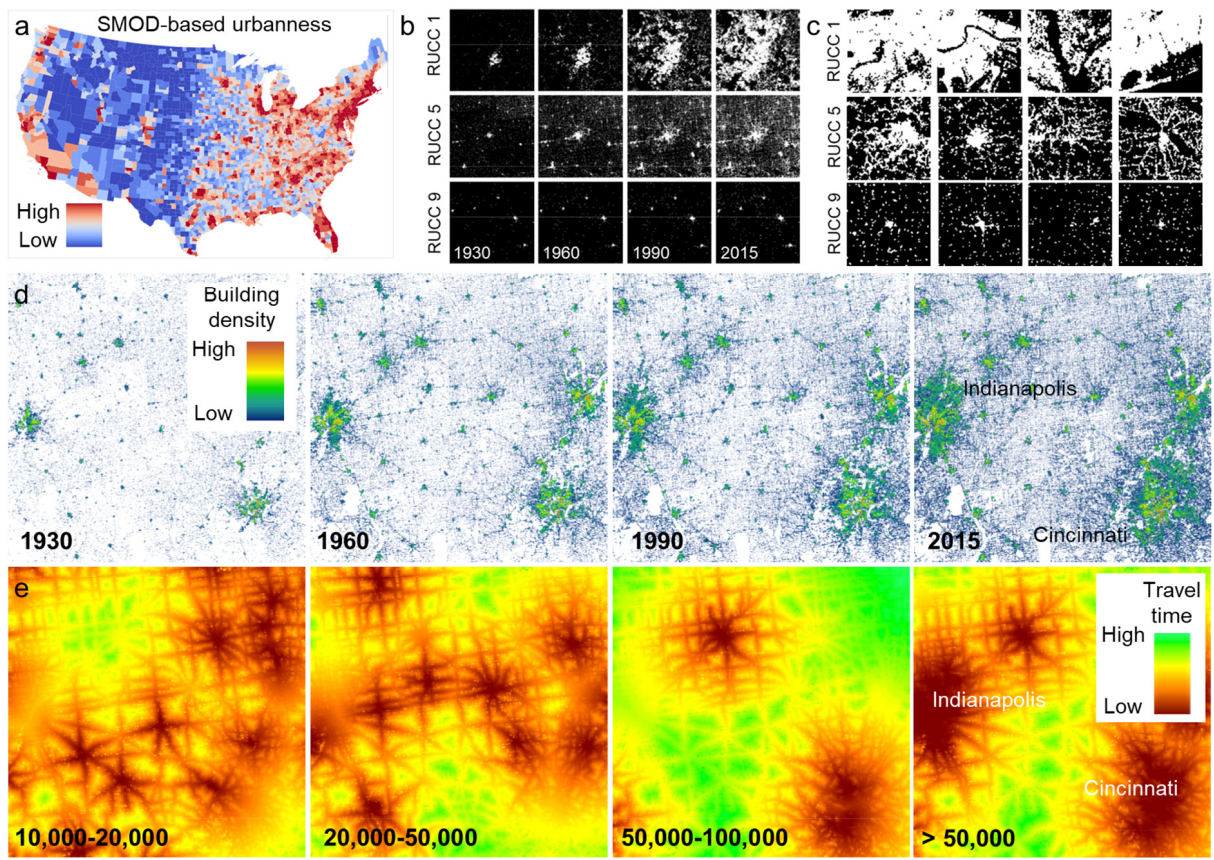

**Figure 6.** Data used for the evaluation of the proposed remoteness indices: (a) GHS-SMOD based county-level urbanness score, (b) HISDAC-US place-level built-up areas 1930 – 2015, (c) MSBF built-up areas, both shown for illustrative examples of places within counties of different RUCC, (d) HISDAC-US building density 1930-2015 shown for the Indianapolis-Cincinnati region, and (e) travel-time based accessibility indicators to cities of different population ranges from Nelson et al. (2019) shown for the same region.

## 3. Results

### 3.1. The rural-urban continuum in the U.S. from 1930 to 2018

Fig. 7 shows the PLURAL indices for the two modelling approaches, and for the equally weighted scenarios, for 1930, 1970, and 2018 (see Fig. A2 for maps of all weighting schemes). Indices are scaled into the range [0,1] jointly across all years, and thus, the obtained indices are comparable over time. Both modelling approaches yield similar broad-scale patterns, and reflect the commonly known settlement patterns in the U.S. during the 20[th] century (i.e., early settlements in the Northeast, and late, fast-growing urban areas in the South). Notably, some differences occur at local scales, particularly in early points in time: While the raster-based approach identifies early settlements such as Tucson (AZ), Santa Fé (NM), Albuquerque (NM) as highly urban (i.e., low remoteness), such extreme local differences are not present in the 1930 network-based result, which appears to be more sensitive to smaller places of local or regional importance (e.g., lower remoteness levels along historical trade and settlement routes such as the Oregon trail). The detailed dynamics of the rural-urban continuum from 1930 to 2018, as modelled by these approaches can be seen in Supplementary Movie 1 and in Supplementary Movie 2 and as individual animations for each of the two indices, for each scaling type (i.e., per year and across all years) and for each weighting scheme (http://doi.org/10.3886/E162941).

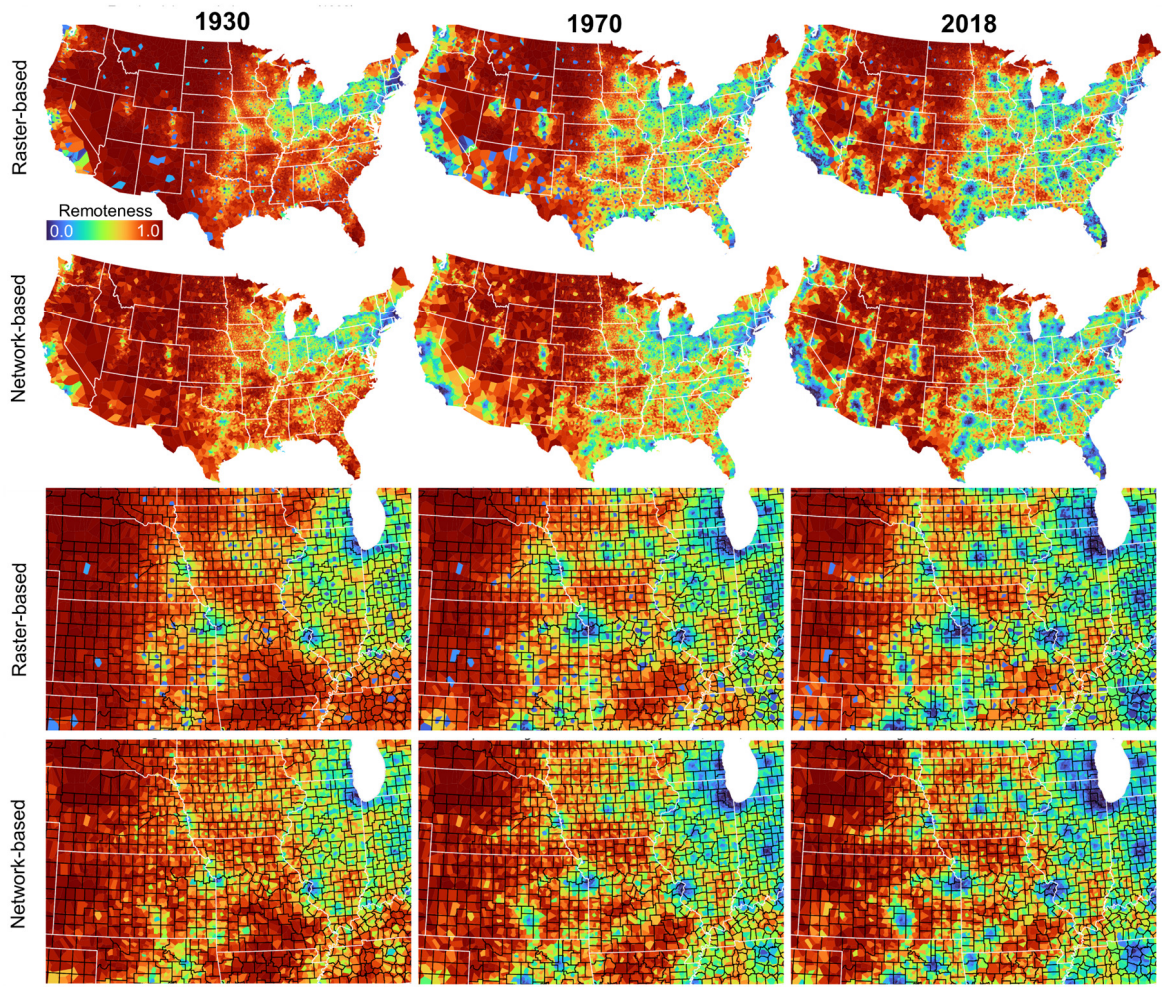

**Figure 7. Equal-weighting schemes for raster-based (PLURAL-1, top row) and network-based (PLURAL-2, bottom row) remoteness indices in 1930, 1970, and 2018, for CONUS (top) and the U.S. Midwest (bottom); indices are scaled jointly across all points in time and thus, comparable over time. In the enlargements, county boundaries are shown in black, and state boundaries in white. In these maps, each place is represented by the Thiessen polygon established from the discrete place locations.**

### 3.2. Comparing the two methods

The quantitative comparison of the two modelling approaches yields high levels of correlation (Pearson > 0.8) between any modelling approach and weighting scheme, and these correlations slightly decrease over time (Fig. 8a,d). The scatterplots of the equal-weights schemes (Fig. 8b,e) indicate that for the large majority of places, the network-based approach yields a more conservative remoteness estimate, i.e., most places are below the main diagonal, thus, PLURAL-2 indicates higher levels of urbanness. Notably, the relationship between the two approaches differs between smaller and larger places (as defined by the place population), exhibiting two clusters, shifted along the x-axis: While for large places, PLURAL-1 and PLURAL-2 yield similar values, for small places, PLURAL-1 yields higher levels of remoteness (approximately +0.2) than PLURAL-2. Moreover, the spatial patterns of the differences between PLURAL-1 and PLURAL-2 estimates per place exhibit a strong spatial pattern in East-West direction, indicating that the previously observed systematic offset of +0.2 is particularly prevalent in the East, whereas places where network-based remoteness exceeds raster-based remoteness, are mostly smaller places in the West, likely due to the higher sensitivity of the network-based approach to regional or local importance of places, and a result of the east-west population density gradient in the CONUS.

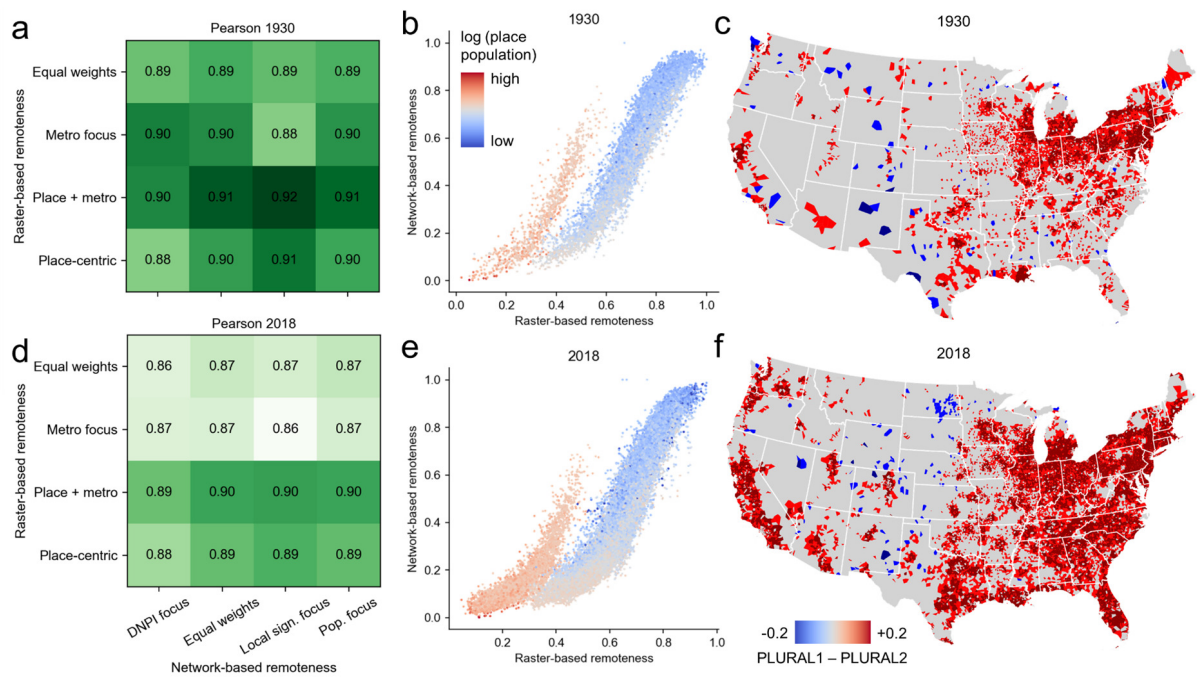

**Figure 8.** Comparison of raster-based and network-based remoteness indices: (a) Pearson's correlation coefficients between the four raster- and network-based weighting schemes, (b) Scatterplot of the equally weighted raster- and network-based remoteness indices in 1930, color-coded by place population, and (c) corresponding map of the differences in remoteness. Panels (d) to (f) show the corresponding results for 2018. In maps (c) and (f), places are represented by the Thiessen polygons established from the discrete place locations.

### 3.3. Comparison to county-level rural-urban classifications and GHS-SMOD

Comparing the place-level remoteness index distribution within strata defined by the county-level urban-rural designations (Section 1), we generally observe a coherence between the PLURAL indices and county-level classes, manifesting in increasing place-level remoteness with increasing county-level rurality, for all county-level classification schemes and for both modelling approaches (Fig. 9a,b). Notably, we observe high levels of "lower" outliers (i.e., below the lower whisker defined as 1.5xIQR) for the raster-based approach (PLURAL-1) (Fig. 9a), located evenly across the CONUS (Fig. 9c). This effect is not present when comparing the network-based remoteness index (PLURAL-2) against the county-level designations (Fig. 9b,d), thus indicating that the spatial-network approach better approximates the rural-urban gradient models underlying the county-level classifications. Such an effect is also observed for the other weighting schemes (Fig. A6). This effect is particularly strong when comparing against the GHS-SMOD based strata (Fig. 9a,b), indicating that the spatial network approach, capable of capturing the spatial configuration of places, is able to approximate a population-density and built-up land-based RUC modeling approach based on population data only. The strong spatial patterns of "upper" outliers, highly similar across the different county-level classifications (Fig. 9c,d) indicate counties of presumably high levels of within-county variability of remoteness, for example in Arizona and North Dakota (Fig. 9c). In other words, the spatial refinement in using place-level remoteness instead of county-level estimates is assumed to be particularly effective in these regions.

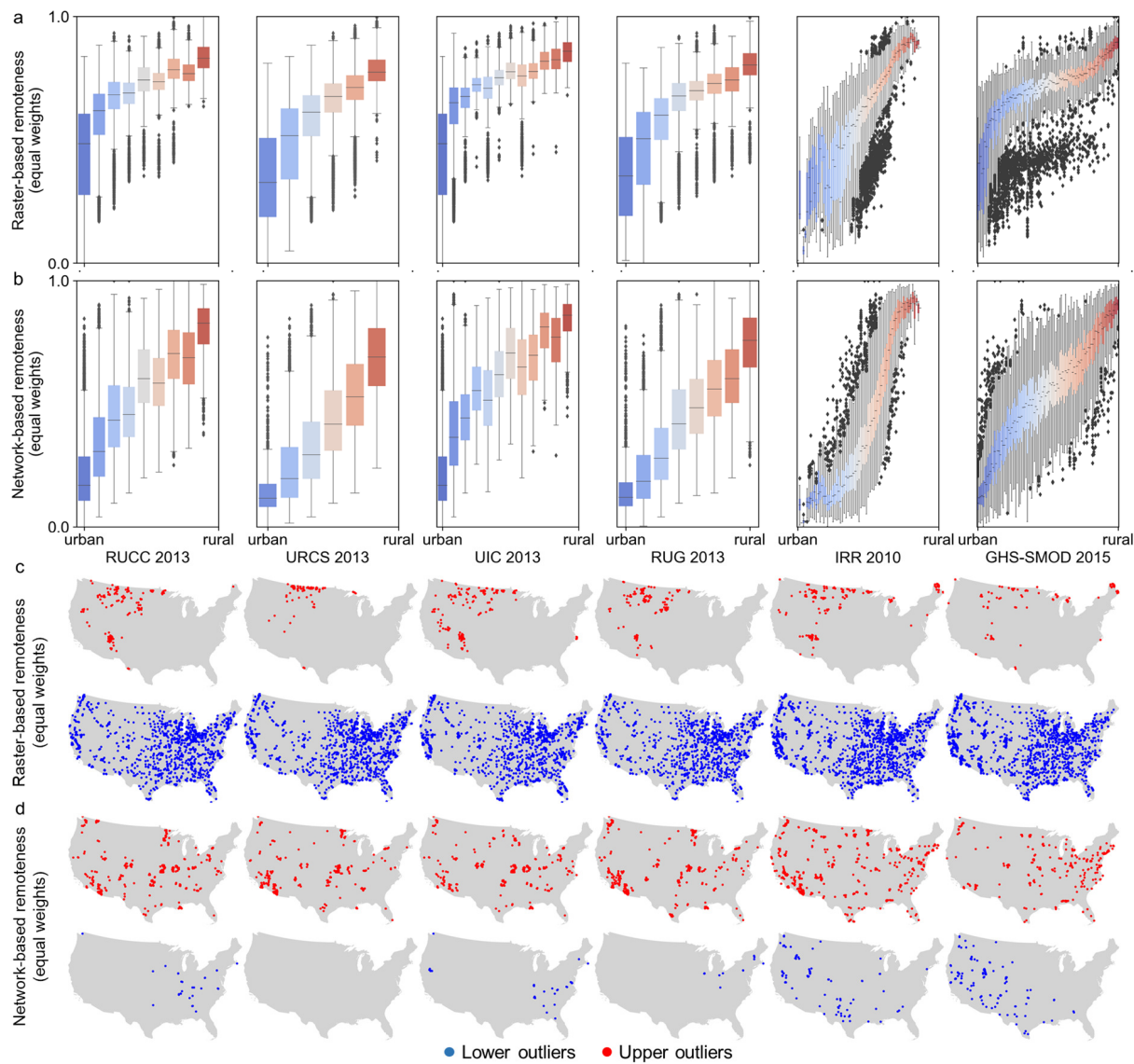

**Figure 9.** Comparison of the place-level remoteness indices and their county-level rural-urban designations. (a) Raster-based remoteness distributions and (b) network-based remoteness distributions in county-level rural-urban classes, both shown for the equal-weights scenario. Panels (c) and (d) show the spatial distributions of the upper and lower outliers indicated in the box-and-whisker plots, for the raster- and network-based remoteness, respectively. Coloring of the boxes in (a) and (b) corresponds to the classes / values of the county-level classifications (blue=urban, red=rural).

Visually, the trends observed in Fig 9a,b indicate varying relationships between the place-level remoteness estimates and county-level classes. We formally tested linearity- and rank-based correlation between the county-level RUC classifications and our place-level indices. While the rank-based Spearman's correlation coefficient is high (>0.7) for all indices and county-level classes (Fig. 10), we observe highest levels of linearity for the network-based approaches and GHS-SMOD, confirming the previous observations.

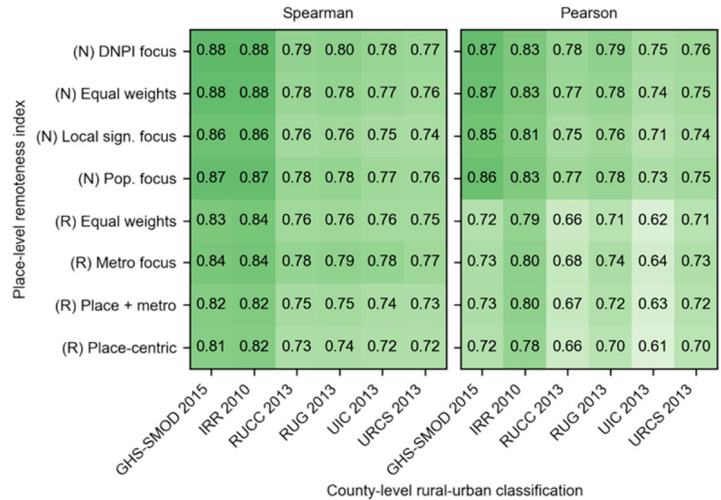

**Figure 10. Spearman and Pearson correlation coefficients between the place-level remoteness indices (R=raster-based, N=network-based) and the corresponding county-level rural-urban designations.**

### 3.4. Comparison to 2010 ZIP-code level FAR remoteness measures

While most of the discussed rural-urban classifications for the U.S. are based on proximity to metropolitan areas, or commuting patterns, the Frontier And Remote area codes (FAR), available at the ZIP code level, is the only index explicitly implementing the concept of remoteness by means of road-network derived travel times, and population sizes (Section 2.4.1). The ROC plots shown in Fig. 11 show generally high Area-under-the-Curve (AUC) values, indicating that for each of our remoteness indices, there is a threshold that allows for mimicking the FAR classes at low levels of type I and type II errors (i.e., False positive rate < 0.2, True positive rate > 0.9). The remaining disagreement may be due to the ambiguous spatial relationship between ZIP code areas and census places, their difference in spatial granularity and different modelling strategies (i.e., use of Euclidean distance versus road network distance). Here, it is worth noting that ZIP code areas outside of place boundaries are not taken into account in this assessment.

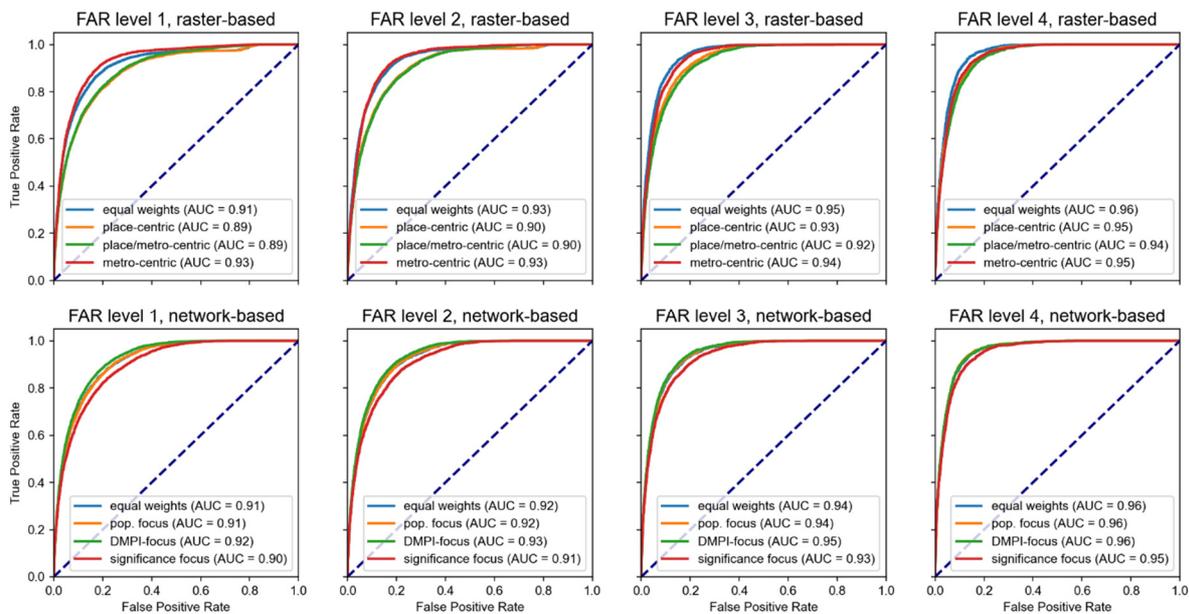

**Figure 11. Receiver-operator-characteristic (ROC) analysis to test how well the place-level remoteness indices can mimic fine-grained, travel time based remoteness classes from the Frontier and Remote area (FAR) codes dataset, shown for FAR classes 1-4, and for each weighting scheme of the raster-based (top row) and network-based approach (bottom row). The dashed line represents the ROC curve of a random relationship between the remoteness indices and FAR class memberships for comparison.**

## 3.5. Long-term place-level trends of the built environment across the rural-urban continuum

When comparing the number of built-up records (i.e., approximate number of buildings) and the built-up area against our remoteness indices, per census place and over time, we observe a highly nuanced relationship between built-environment characteristics and the rural-urban continuum as modelled by our remoteness indices. As expected, high levels of remoteness are associated with low building counts and small built-up areas. These patterns are highly stable over time, and across weighting schemes when visually assessing the scatterplots in Fig. 12a,b. These point patters differ slightly between the raster-based approach and the network-based approach, indicating that the ability to model a built environment perspective of rurality based on population data varies between approaches. Moreover, the negative correlations between building counts / built-up area and remoteness increase over time (Fig. 12c,d), in particular for the built-up records variable (Fig. 12c). While the discussion of the drivers for these changes over time is out of the scope of this paper, a possible reason could be lower levels of completeness in the HISDAC-US and underlying ZTRAX data in early points in time (Uhl et al. 2021).

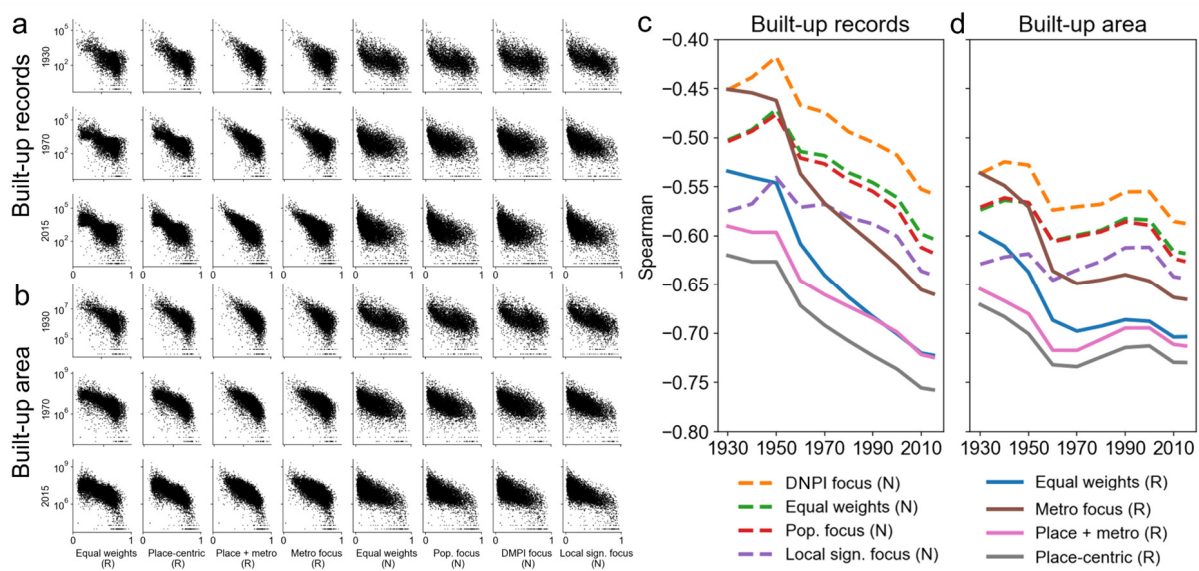

**Figure 12. Relationships of ZTRAX-based built environment characteristics and the proposed remoteness indices based on the raster approach (R) and the network approach (N) over time. (a) Scatterplots of place-level remoteness and building counts, and (b) built-up area in 1930, 1970, and 2015, respectively; Panel (c) and (d) show corresponding time series of Spearman's correlation coefficient. Note that 2015 built environment characteristics are compared to the 2018 remoteness indices.**

## 3.6. Relationship of remoteness to contemporary landscape and settlement characteristics

The assessment of the relationship between size and structure related characteristics of the contemporary built-up land (derived from Microsoft's U.S. building footprint data) at the place level reveals high levels of association of these characteristics with the place-level remoteness, which is in line with related work using landscape metrics as a proxy for the rural-urban gradient (Luck & Wu 2002, Vizzari 2011, Vizarri 2013). We generally observe high levels of negative correlation between remoteness and the settlement and landscape characteristics, indicating that remote places are characterized by few, small buildings, organized in small, highly disconnected patches of built-up area of similar sizes (Fig. 13a). These negative correlations are slightly higher for the network-based modelling approach than for the raster-based remoteness indices (Fig. 13a), and the patterns of these measures across the RUC are highly similar for the different weighting schemes (Fig. 13b). These findings confirm once more that our remoteness models are consistent with the literature and exhibit an expected behavior when compared to landscape metrics.

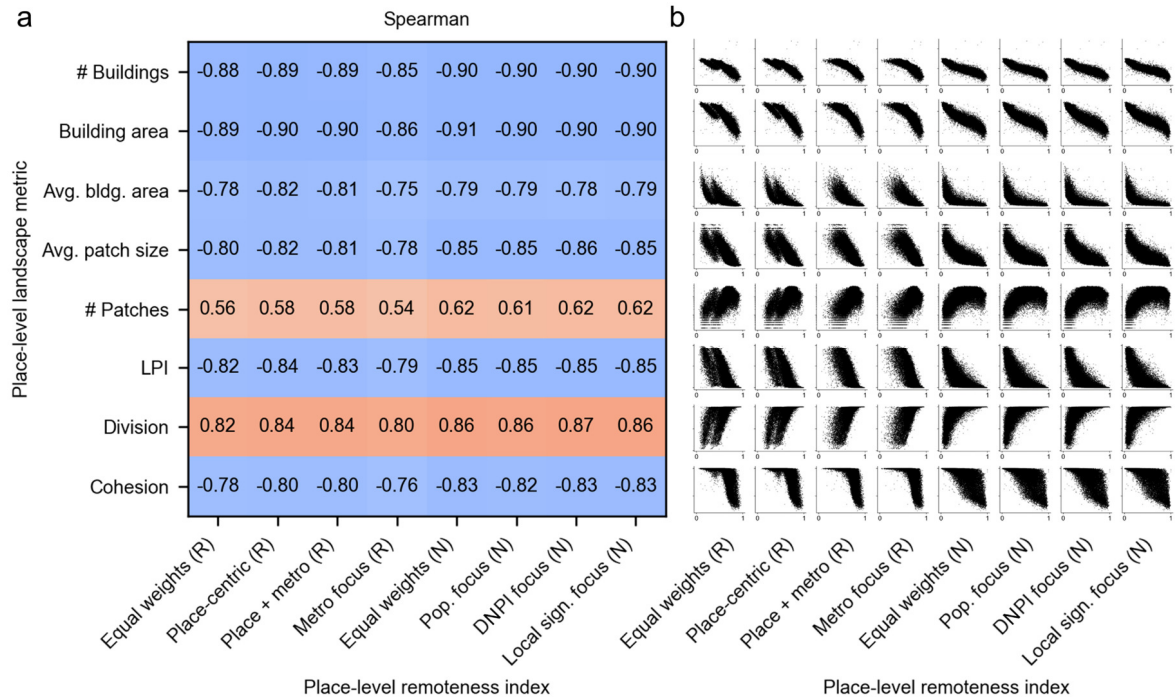

**Figure 13. Assessing the relationships between the remoteness indices and contemporary building and landscape metrics derived from Microsoft building footprint data. (a) Spearman's rank correlation, and (b) corresponding scatterplots.**

### 3.7. Comparison to travel-time accessibility indicators

Lastly, we assess the impact of using Euclidean distance instead of road network distance to model remoteness, using the travel times from Nelson et al. (2019) observed at each 2018 place centroid. We generally observe increasing travel times with increasing remoteness, while this trend seems to be slightly more linear for the network-based indices than for the raster-based indices, as the scatterplots in Fig. 14a suggest. For the raster-based equal-weights and place-centric weighting schemes, we observe an additional peak in travel times for remoteness values around 0.4, in particular for the travel times to places >50,000 inhabitants, as an effect of the local focus of these weighting schemes. Comparing travel times to the corresponding Euclidean distances $D_{PC}$ (see Eq. 1) used for the raster-based remoteness models, we observe a noisy, but roughly linear relationship (Fig. 14b). Places that deviate heavily from the main diagonal are expected to be located nearby geographic obstacles such as mountain ranges, rivers, borders, or lakes, causing deviations of road network routes from the shortest (Euclidean) route (Fig. 14b). The visualization of the theoretical travel speed estimates obtained from the Euclidean distances and travel time estimates (Fig. 14c) shows that for most parts of the CONUS, these travel speed measures range within plausible values, i.e., between 30 km/h and 140 km/h. We derived these thresholds from the typical speed limits (25 miles/hour in cities, 80 miles/hour on interstates) plus approximately 10 km/h of tolerance to account for differences in the definition of places in our modelling approaches and the Nelson et al. (2019) accessibility indicators. For example, Nelson et al. seem to model Chicago as a single place, whereas U.S. census places distinguish between different neighborhoods of the city. Only in a few regions (in black, and yellow, respectively) this range of plausible travel speed is exceeded. This indicates that for most parts of CONUS, the bias introduced by using Euclidean distance instead of road network distance is within acceptable margins.

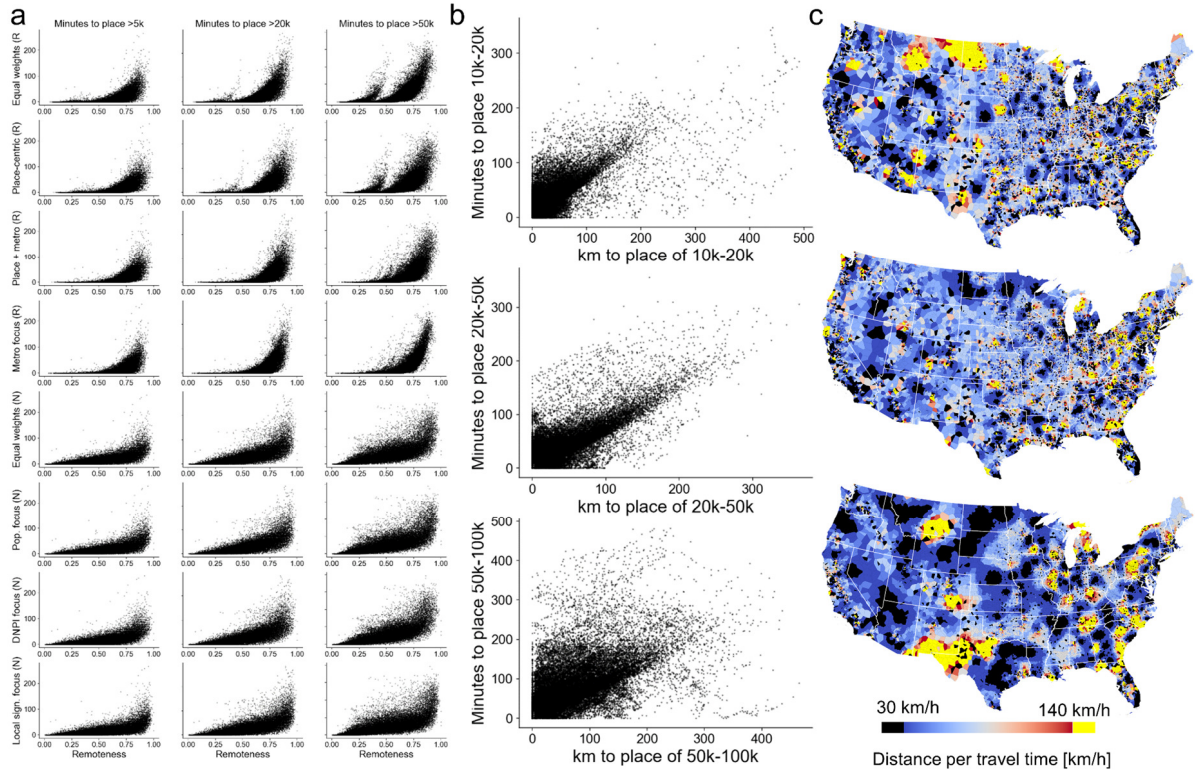

**Figure 14.** Comparing the remoteness indices and their components to travel-time based accessibility indicators (Nelson et al. 2019). (a) Relationships between the different weighting schemes of the proposed raster- and network-based remoteness indices and travel times to the nearest city of population >5k, > 20k, and >50k, respectively, (b) relationships between Euclidean distance based components of the raster-based remoteness approach and travel times, and (c) place-level plausibility analysis of Euclidean distance and travel time by visualizing theoretical travel speed. In (c), places are represented by the Thiessen polygons established from the discrete place locations.

## 4. Discussion and conclusions

Herein, we described and evaluated two approaches to model the rural-urban continuum at fine spatial grain (i.e., the place level) and over long time periods (i.e., 1930-2018) for the conterminous US. This work makes several contributions: (a) We fill a gap in the US data landscape by providing temporally consistent, place-level rural-urban classifications, refining existing county-level classifications, and complementing finer-grained classifications at the ZIP code level by providing a demographically meaningful analytical unit (i.e., the census place), applicable over very long time periods, (b) We generate methodological knowledge by developing and comparing a raster-based approach and a spatial network approach. (c) Specifically, we adopt elements of a spatial network approach developed from a physical perspective on human settlements (Esch et al. 2014) to a population perspective of human settlements; Moreover, we introduce a novel and effective metric for measuring remoteness based on population distributions discretized to point data, which we call the ***distance-based neighborhood population index*** (***DNPI***). (d) We make all of our place-level remoteness indices publicly available, calling them the ***place-level urban-rural index*** (***PLURAL***), enabling researchers of various disciplines to conduct fine-grained, cross-sectional and longitudinal analyses across the rural-urban continuum, and over a time period of almost 90 years.

The cross-comparison of the created indices against each other, and against five external datasets of different nature, revealed that our indices, despite being based on simple data structures, effectively model the rural-urban gradient patterns. Moreover, our analyses have shown that the network-based remoteness index (PLURAL-2) seems to exhibit higher levels of coherence to the data used for comparison than the raster-based approach (PLURAL-1). However, these differences largely play out at the urban side of the rural-urban continuum, i.e., mostly affecting large places. Table 2 summarizes the major findings from each of the performed evaluation analyses.

**Table 2.** A brief summary of the different evaluation studies of the PLURAL indices between each other and against external datasets.

| Assessment | Section | Major finding |
|---|---|---|
| Comparing PLURAL-1 to PLURAL-2 | 3.2. | The network-based approach (PLURAL-2) tends to classify places as more urban than the raster-based approach (PLURAL-1). |
| Comparing to county-level classifications | 3.3. | PLURAL-2 exhibits far fewer outliers when analyzing distributions within county-level groups; PLURAL-2 appears to agree more with county-level classifications. |
| Comparing to FAR codes | 3.4. | All indices show high agreement with travel-time based FAR classes, indicating that the Euclidean distance-based approximation does not meaningfully bias the resulting indices. |
| Comparing to long-term built-environment trajectories | 3.5. | We observe a negative correlation between remoteness and total buildings per place. This negative correlation increases over time, indicating greater plausibility in our indices for more recent points in time. |
| Comparing to contemporary landscape metrics | 3.6. | We observe strong associations between remoteness and morphological characteristics (landscape metrics) for all indices and weighting schemes. |
| Comparing to travel-time based accessibility indicators | 3.7. | The implemented distance concept using Euclidean distance rather than road network distance introduces low levels of bias overall but may distort the results for a small number of individual places. |

The indices and weighting schemes have strengths in different areas and we recommend the following: For applications of the PLURAL indices *identifying rural places*, both methods seem to work well. For studies *across the whole urban-rural continuum*, we recommend practitioners to use PLURAL-2, as it is more normally distributed across the RUC (see Figs. 9, A6). For *longitudinal studies*, the PLURAL-1 (scaled across all years) is more suitable, since the network-based PLURAL-2 contains a rank-based component (see Section 2.3.2) and thus, is not fully comparable over time. The choice of the weighting scheme depends on the individual application.

The PLURAL indices will enable researchers to work at spatial units that are potentially more meaningful to certain rural processes such as the role of place characteristics in social mobility (Chetty et al. 2014; Connor & Storper 2020), health (Manduca & Sampson 2019; Shah et al. 2020), and voting (Sachdeva et al. 2021). Such refined scales can be also used to gain novel insight into the spatial distributions of social vulnerability (Spielman et al. 2020), public health issues (Anderson et al. 2021), or the exposure to natural hazards risks (Braswell et al. 2021). Moreover, the presented indices cover a long time period, 1930 to 2018, and are fully consistent over time, enabling longitudinal analyses of long-term, dynamic processes along the rural-urban continuum.

While this constitutes a useful addition for many applications, our approach still faces some notable limitations. For example, the raster-based approaches are based on population density calculated within focal windows of a fixed radius, and based on commonly used population categories. We will explore the sensitivity of our indices to the choice of focal window size and population thresholds in future work. Moreover, the PLURAL indices do not take into account territory and populations outside of census places, and thus, they cannot be used for assessments of spatial processes taking place in unincorporated land outside of census place boundaries. As noted above, however, we contend that proximate places are of relevance to non-place-based populations as a hub of local social and economic activity. From this, we infer that the level of rurality assigned to a place is usually a strong reflection of the rurality of surrounding areas. Moreover, in order to apply these methods to other countries, border effects need to be taken into account, given the highly developed levels of cross-country mobility (e.g., in Europe).

Hence, future work will include the generation of continuous, local remoteness measures, e.g., at the grid cell level, to create a fine-grained, gapless model of the rural-urban continuum. The Thiessen polygons that we used to visualize places (Fig. 7) and to generate spatial networks (Fig. 4b) could be further combined to model the areal influence of places, potentially enabling us to allocate non-place populations to proximate places. Future work will also include the application of the presented approaches to other countries where comparable data is available, or even to expand such efforts to continental scales. Concluding, it is our hope that the PLURAL remoteness indices will enable researchers to add a long-term temporal dimension to rural and rural-urban studies, at a refined spatial granularity, and ultimately, contribute to more informed planning and decision-making.

## 5. Data availability

The PLURAL indices and the underlying historical place-level population counts, as well as the derived remoteness indicators used to establish the PLURAL-1 and PLURAL-2 indices are available as tabular data (CSV format). Moreover, separate spatial vector data files containing the place locations for each point in time (1930-2018) attributed with the PLURAL indices are available in ESRI Shapefile format at http://doi.org/10.3886/E162941.

## 6. Supplementary materials

We created animated visualizations illustrating the dynamics of the rural-urban continuum in the CONUS for the time period from 1930 to 2018. These data animations include (a) a comparison of the four equally-weighted indices for the CONUS (Supplementary Movie 1), and (b) an enlarged version for the Midwest (Supplementary Movie 2), as well as individual animations for each of the indices, for each scaling type, and for each weighting scheme (see http://doi.org/10.3886/E162941).

**Acknowledgments.** This research has been supported by Project Number R21HD098717-01A1, "Health, Social, and Demographic Trends in Rural Communities" funded by the Eunice Kennedy Shriver National Institute of Child Health and Human Development (NICHD). The project has also benefited from research, administrative, and computing support, also provided by NICHD, to the University of Colorado Population Center (CUPC; Project 2P2CHD066613-06). The content is solely the responsibility of the author and does not necessarily represent the official view of the CUPC, National Institutes of Health (NIH) or CU Boulder. Furthermore, the authors would like to thank the student workers Brenda Blanco Soto and Ciara Coughlan for their help with digitizing historical census records used in this study.

**Funding.** Funding for this research was provided by the University of Colorado Population Center (Project 2P2CHD066613-06), funded by the Eunice Kennedy Shriver National Institute of Child Health and Human Development. The content is solely the responsibility of the author and does not necessarily represent the official views of the CUPC, NIH or CU Boulder.

**Author Contributions.** J.U. and L.H. designed the indices. J.U. and S.L. designed the evaluation analysis. C.T., C.H., M.G., and D.C. gathered census data. J.U. processed the data and gathered evaluation data. J.U. conducted the evaluation experiments and visualized the results. J.U., L.H., S.L., D.C., J.N., C.H., and M.G. wrote the paper.

**Declarations of interest:** None.

## 8. Appendix

**Table A1. Weighting schemes for raster-based remoteness index.**

| Weighting scheme | Place population | Focal population density | Distance (10k-20k) | Distance (20k-50k) | Distance (50k-100k) | Distance (100k-250k) | Distance (>250k) |
|---|---|---|---|---|---|---|---|
| Equal weights | 0.143 | 0.143 | 0.143 | 0.143 | 0.143 | 0.143 | 0.143 |
| Place-centric | 0.250 | 0.250 | 0.100 | 0.100 | 0.100 | 0.100 | 0.100 |
| Place-centric & metro focus | 0.250 | 0.250 | 0.033 | 0.067 | 0.100 | 0.133 | 0.167 |
| Metro focus | 0.100 | 0.100 | 0.053 | 0.107 | 0.160 | 0.213 | 0.267 |

**Table A2. Weighting schemes for network-based remoteness index.**

| Weighting scheme | $POP_{Place}$ | $NPD_1$ | $NPD_2$ | $NPD_3$ | $MLS_1$ | $MLS_2$ | $MLS_3$ | $DNPI_{C3}$ | $DNPI_{250km,500k}$ | $DNPI_{500km,1,000k}$ | $DNPI_{MAXPOP}$ |
|---|---|---|---|---|---|---|---|---|---|---|---|
| Equal weights | 0.091 | 0.091 | 0.091 | 0.091 | 0.091 | 0.091 | 0.091 | 0.091 | 0.091 | 0.091 | 0.091 |
| Population focus | 0.125 | 0.125 | 0.125 | 0.125 | 0.071 | 0.071 | 0.071 | 0.071 | 0.071 | 0.071 | 0.071 |
| DNPI focus | 0.071 | 0.071 | 0.071 | 0.071 | 0.125 | 0.125 | 0.125 | 0.125 | 0.071 | 0.071 | 0.071 |
| Significance focus | 0.071 | 0.071 | 0.071 | 0.071 | 0.071 | 0.071 | 0.071 | 0.071 | 0.167 | 0.167 | 0.167 |

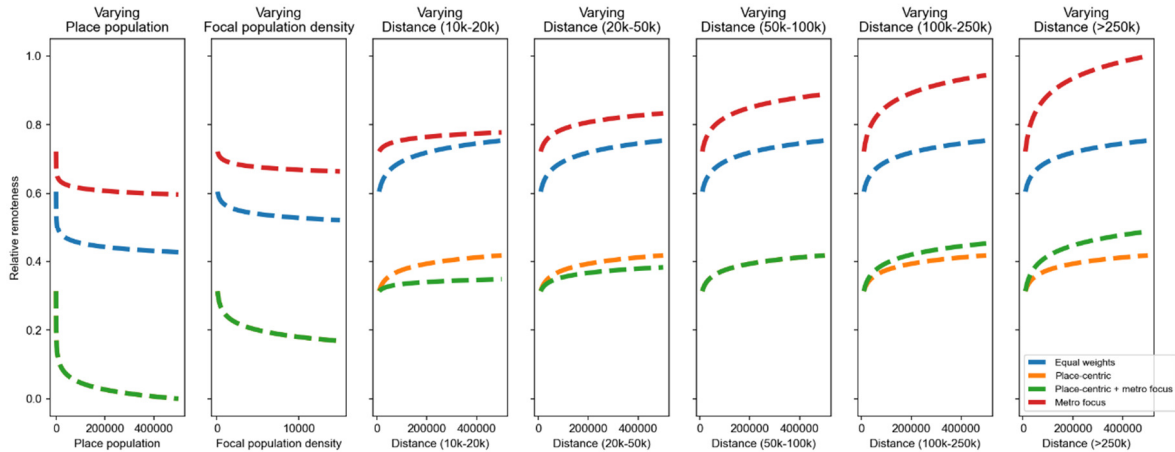

**Figure A1. Sensitivity of different weighting schemes to changes in individual remoteness indicators for the raster-based approach. A synthetic baseline place with randomly initialized remoteness indicator was used, and each of the seven indicators was systematically increased. For each increment, the resulting remoteness index was calculated, for each of the four weighting schemes. The curves show the different levels of sensitivity of the weighting schemes.**

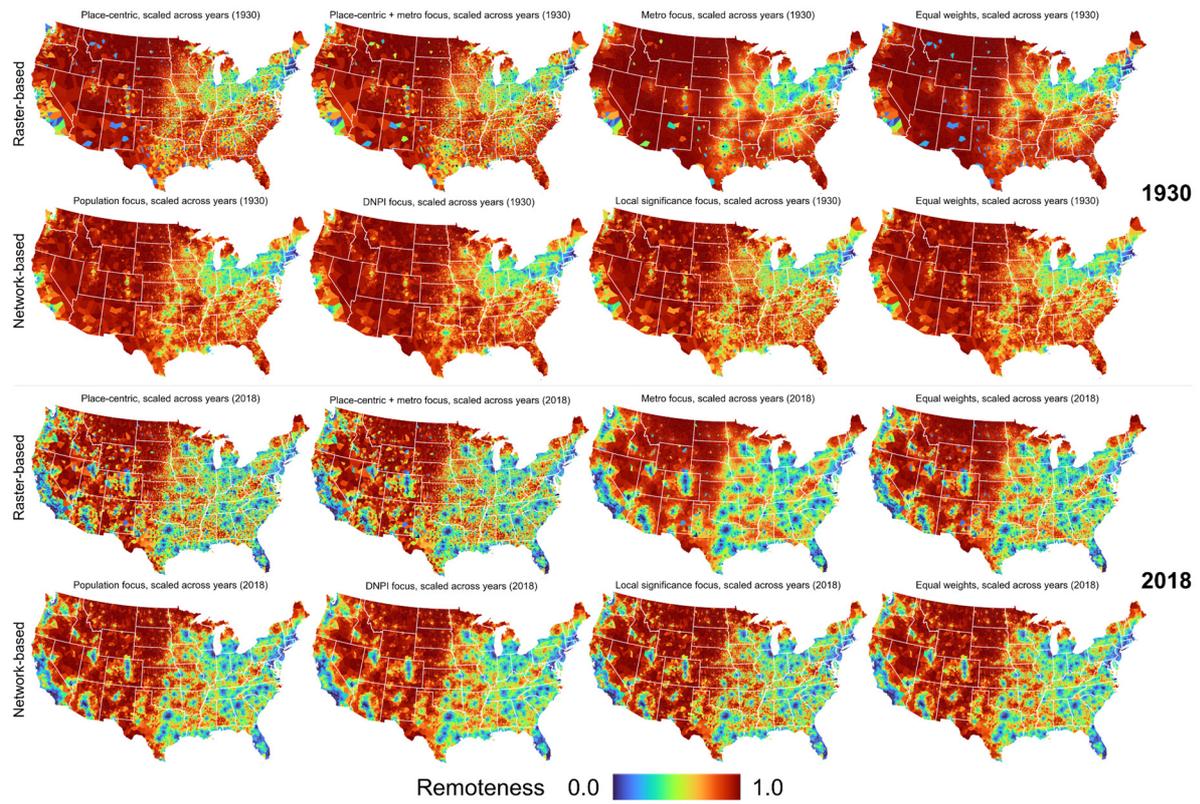

Fig. A2. Maps of the remoteness indices in 1930 (top) and 2018 (bottom), each shown for the raster-based approach and network-bases approach, and for each of the four weighting schemes. In these maps, each place is represented by the Thiessen polygons established from the discrete place locations.

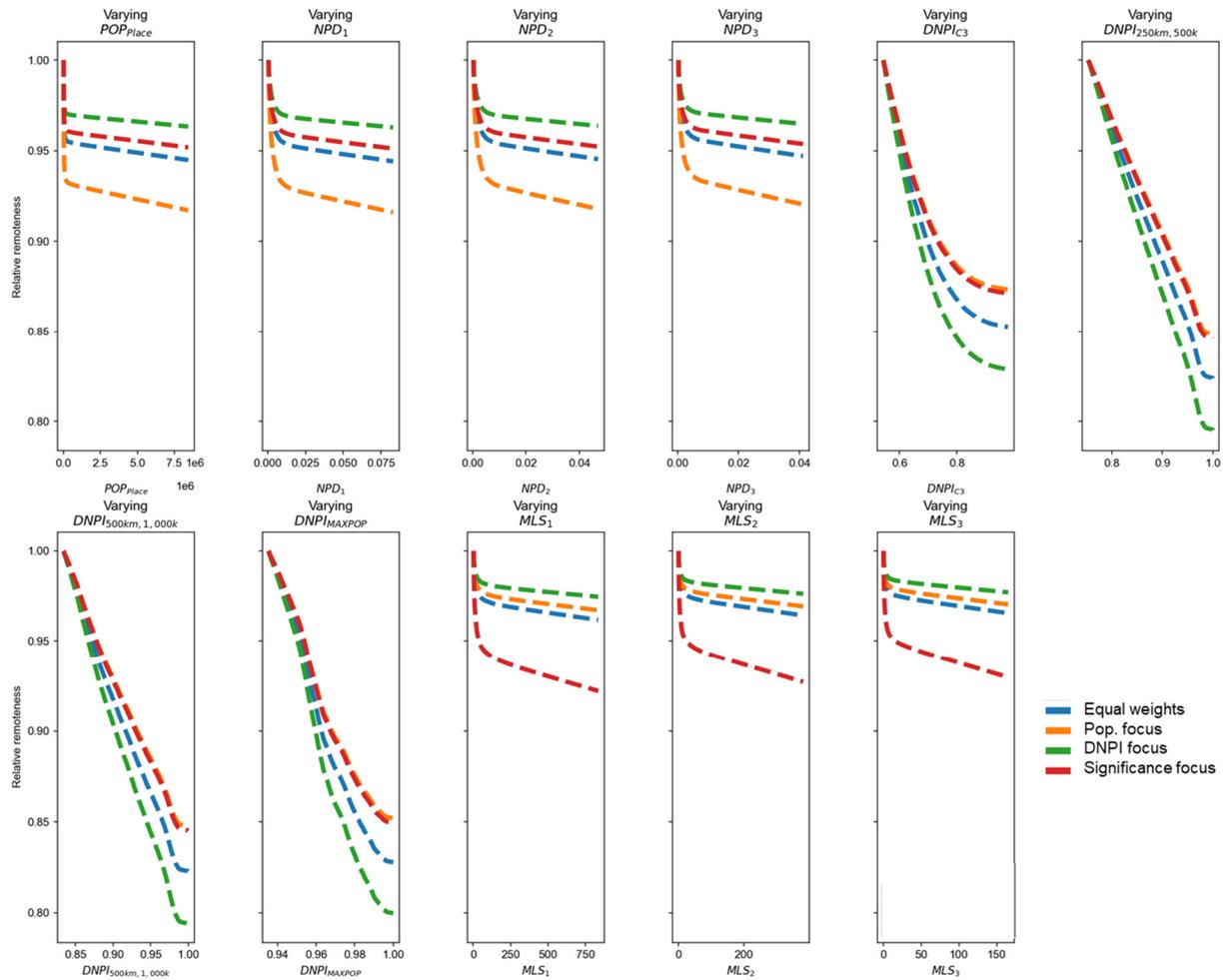

**Figure A3. Sensitivity of different weighting schemes to changes in individual remoteness indicators for the network-based approach.** A synthetic baseline place with randomly initialized remoteness indicator was used, and each of the seven indicators was systematically increased. For each increment, the resulting remoteness index was calculated, for each of the four weighting schemes. The curves show the different levels of sensitivity of the weighting schemes.

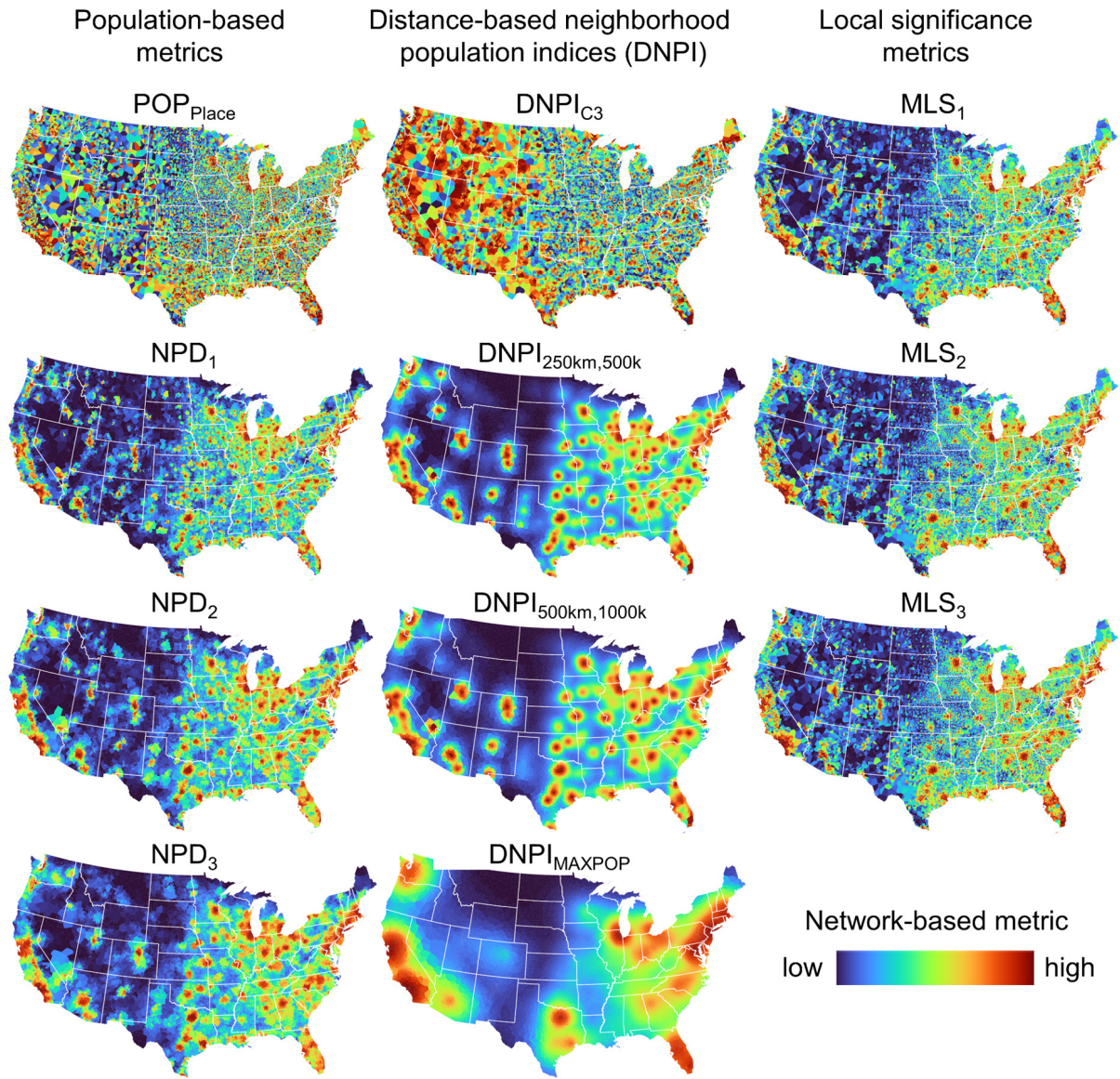

Fig. A4. Maps of the 11 network-based metrics in for each census place in 2018, input to the spatial-network-based PLURAL-2 remoteness indices. Abbreviations: NPD = Neighborhood population density, DNPI = Distnnce-based neighborhood population index, MLS = Median local significance. In these maps, each place is represented by the Thiessen polygons established from the discrete place locations.

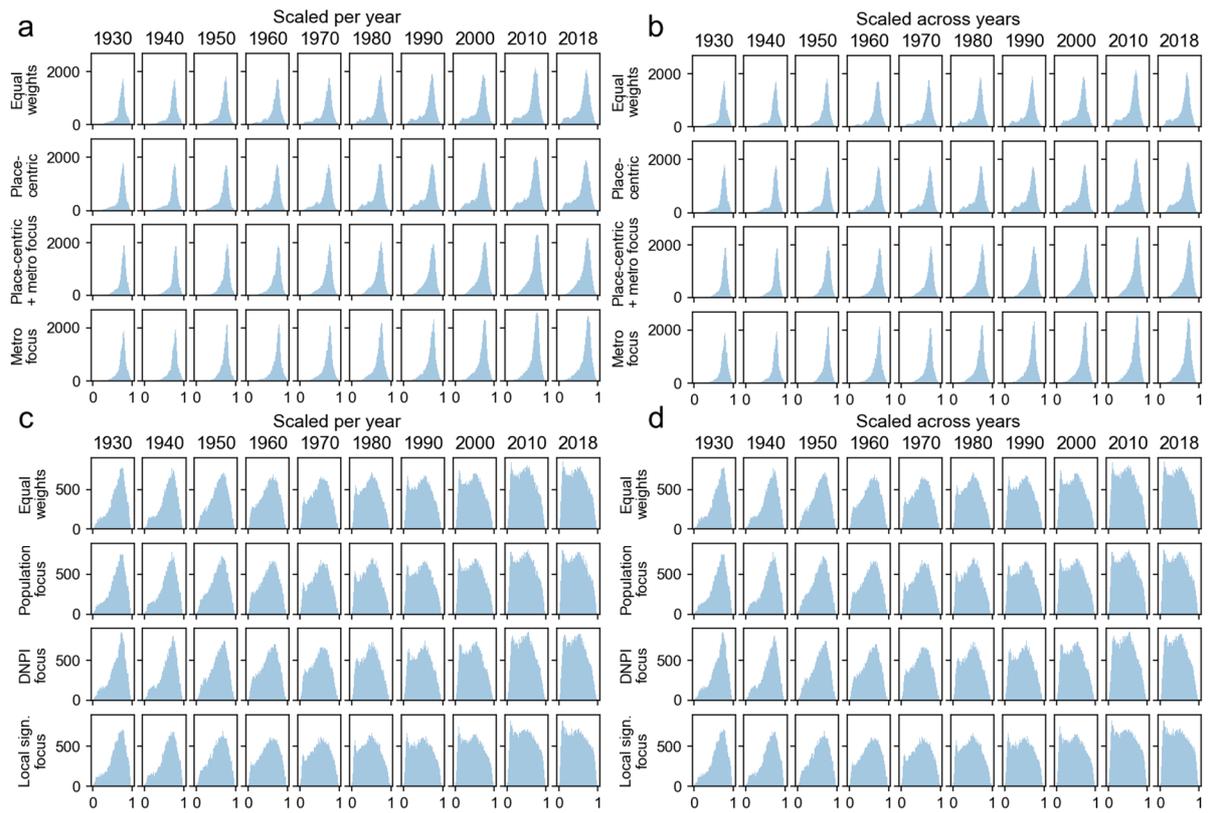

**Fig. A5.** Distributions of remoteness indices for the US 1930-2018 for different weighting schemes and scaling strategies: (a) Raster-based approach (PLURAL-1) scaled per year and (b) scaled across all years, (c) spatial network based approach (PLURAL-2) scaled per year and (d) scaled across all years.

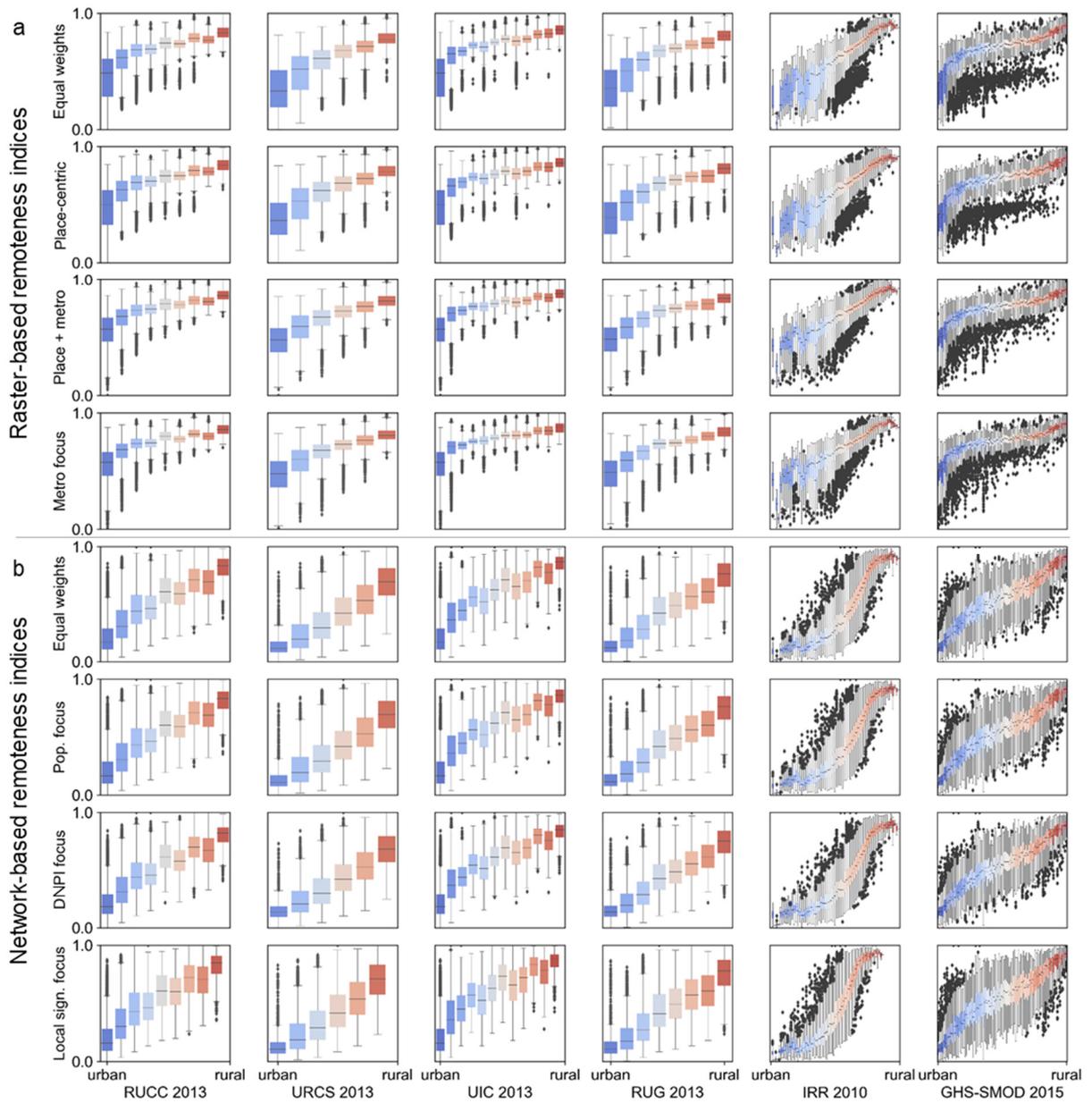

Fig. A6. Comparison of county-level rural-urban classifications and the place-level remoteness indices for each weighting scheme of the (a) raster-based approach, and (b) the network-based approach. Color of boxes represent the individual classes / values of the county-level classifications (blue=urban, red=rural).